\documentclass{aa}  

\usepackage{graphicx}
\usepackage{txfonts}
\usepackage{amsmath,amsfonts,amssymb}
\usepackage{textcomp,gensymb}
\usepackage[T1]{fontenc}
\usepackage{aecompl}
\usepackage{ulem}
\usepackage{aas_macros}
\usepackage[dvipsnames]{xcolor}
\usepackage{color}
\usepackage{comment}
\usepackage{times}
\usepackage{subfig}
\usepackage{stfloats}
\usepackage{subcaption}
\usepackage[utf8]{inputenc}
\usepackage{xcolor}
\usepackage[
  colorlinks=true,
  linkcolor=blue,
  citecolor=blue,
  urlcolor=blue
]{hyperref}
\usepackage{multirow}
\usepackage{natbib}
\usepackage{placeins}
\usepackage[percent]{overpic}

\begin{document} 

   \title{Stability of circumbinary planets: the role of binary properties and migration scenarios}

   \author{Romain Grane\inst{1}\fnmsep\thanks{email adress: romain.grane@univ-grenoble-alpes.fr}
          \and
          Nicol\'as Cuello\inst{1}
          \and
          Mario Sucerquia\inst{1}
          \and
          Emmanuel Gianuzzi\inst{2}
          \and
          Milenne Ávila-Bravo\inst{3}
          \and\\
          Carolina Charalambous\inst{3}
          \and
          Amaury H. M. J. Triaud\inst{4}
          }

   \institute{Univ. Grenoble Alpes, CNRS, IPAG, 38000 Grenoble, France
             \and
             Instituto de Astronomía Teórica y Experimental (IATE), UNC-CONICET, Laprida 854, (X5000BGR) Córdoba, Argentina
             \and
             Instituto de Astrofísica, Pontificia Universidad Católica de Chile, Av. Vicuña Mackenna 4860, 782-0436 Macul, Santiago, Chile
             \and
             School of Physics \& Astronomy, University of Birmingham, Edgbaston, Birmingham B15 2TT, UK
             }

   \date{Received: 1 May 2026 / Accepted: 2 August 2026}
 
  \abstract
   {Among the thousands of exoplanets detected to date, only a very small fraction are classified as circumbinary planets. When considering only low-mass planets, this number becomes negligible. This rarity may be partly explained by observational biases, but also by the challenging dynamical environment that often prevents the long-term stability of planetary orbits in binary systems.}
   {It is therefore essential to investigate the conditions that enable the formation and survival of stable circumbinary planets, with particular attention to the case of low-mass planets.}
   {To this end, we performed N-body simulations coupled with a planet migration prescription to study the post-formation dynamics of two circumbinary planets of 10 Earth masses in a protoplanetary disc. This approach allows us to estimate whether, for a given set of binary and planetary parameters, planets remain bound to their stars or become unstable and ejected.}
   {Based on more than 1,000 simulations, our results indicate that two-planet inward migration can, in some cases, be halted through resonance capture with the binary. In contrast to some previous purely N-body studies, such configurations appear to remain stable over long timescales. We further find that the binary parameters $q_B$ and $e_B$ play a key role in shaping the stability landscape of multi-planet systems. Within the explored parameter space, some regions support long-term stability, whereas others appear highly unstable. These stability regions are also highly sensitive to the migration timescales (i.e., to disk properties and planetary masses). Finally, our simulations suggest that systems in which two planets enter resonance while migrating together are more likely to form stable multi-planet configurations, particularly around highly eccentric binaries.}
   {This constitutes a first step toward identifying the most promising binary star systems likely to host multiple low-mass circumbinary planets. They may serve as valuable guidelines for future observational campaigns, such as those of PLATO and Gaia.}


   \maketitle
   \nolinenumbers

\section{Introduction}
\label{sec:Introduction}

Stellar multiplicity is recognized as a significant factor in the formation and evolution of stars \citep{Offner+2023}. However, the great majority of exoplanets detected to date are found in single-star systems. As of 9 February 2026, the Exoplanet Encyclopaedia\footnote{Available at \url{https://exoplanet.eu/}} lists a total of 7,957 exoplanets and candidates, of which 1,133 ($\approx14\%$) are found in multiple-star systems. These planets can be classified into two groups: S-type planets, which orbit one component of the stellar system, and P-type planets (also called circumbinary planets, or CBPs), which orbit the barycentre of a binary star \citep{Marzari_Thebault_2019}.

Among the population of planets in multiple-star systems, only 17 confirmed circumbinary planets orbit main-sequence binaries. Most of these planets have been detected through transit observations, with the first discovery being Kepler-16 \citep{Doyle+2011}. Two were discovered using the radial-velocity method \citep{Standing+2023,Baycroft+2025}, and one additional planet was identified via eclipse timing variations \citep{Goldberg+2023}. These planets are found around very close binaries (binary separation < 0.25 AU) and tend to have short orbital periods, with 15 out of 17 having periods shorter than one year. However, circumbinary planets are intrinsically difficult to detect, as they must orbit with periods longer than that of their host binary. As a result, they are preferentially found around close binaries, which themselves represent only a small fraction of Sun-like binaries (see Fig. 4 in \cite{Offner+2023}). This effect introduces strong observational biases that may explain the small number of detections. Despite this limited sample, circumbinary planets may be intrinsically common \citep{Armstrong_2024,Martin_Triaud2014}.

Regarding the other characteristics of these planets, almost all planets have been found close to the same plane as the binary motion plane \citep{Penzlin+2021}. Since circumbinary planets are thought to have formed aligned with their host close binary \citep{Foucart_Lai_2013}, eclipsing binaries are preferentially targeted when searching for circumbinary planets. This strongly favours systems in which the planetary and binary orbital planes are already nearly coincident, introducing a significant observational bias. Apart from two, all planets have masses greater than 10 $M_{\oplus}$. The binary eccentricities range from 0.023 to 0.521, and the binary mass ratio, typically defined as the ratio of the mass of the less massive component to that of the more massive one, can take values between 0.2 and 1.0. Most of these planets are located close to the dynamical limit of the binary system \citep{Holman_Wiegert1999}, although some systems, like Kepler-1647 \citep{Kostov_2016} or BEBOP-3 \citep{Baycroft+2025}, have their innermost planet detected far from the binary star. Interestingly, planet formation is expected to be highly unlikely in the innermost region around the binary \citep{Moriwaki_Nakagawa_2004}.

Indeed, it is often assumed that P-type planets form in the same fashion as planets around single stars, that is to say through accretion processes within protoplanetary discs \citep{Armitage_2010}. However, \cite{Meschiari2012} and \cite{Paardekooper+2012} demonstrated that planetesimals in the inner region undergo strong perturbations from the binary which leads to high eccentricities and orbital crossings, preventing planetesimal accretion. Moreover, \cite{Pierens+2020} showed that it is also strongly unlikely to form planets by combination of streaming instability and pebble accretion in the inner regions, due to turbulence which decreases significantly the efficiency of pebble accretion. To explain the current locations of CBPs, the main hypothesis is that they form in the outer regions of the protoplanetary disc and subsequently migrate inward.

This scenario has been extensively studied over the past two decades \citep{Pierens_Nelson2008a,Pierens_Nelson2008b,Kley_Haghighipour2014,Kley+2019}, indicating that planets should halt their migration near the disc inner edge, close to the dynamical limit (also known or referred to as the critical semi-major axis $a_c$). A planet embedded in a protoplanetary disc experiences both Lindblad and corotation torques, and the net torque is generally negative, driving inward migration \citep{Tanaka_2002}. At the inner edge of the disc, however, the positive surface density gradient can balance the net torque, thereby halting the planet’s inward migration \citep{Pierens_Nelson_2007}. The final orbital location of a planet therefore depends primarily on its mass and on the properties of the disc \citep{Pierens_Nelson2013,Mutter+2017,Thun_Kley2018,Penzlin+2021}.
    
Another way to stop planet migration is a first-degree N:1 mean-motion resonance (MMR) between the planet and the binary star. A MMR occurs when two bodies satisfy a near-commensurability between their mean motions, such that specific linear combinations of their orbital frequencies approach zero. This leads to the libration of these linear combinations, which define the resonant angles \citep{Peale_1976}. Such MMRs between circumbinary planets and their host binaries have been observed in hydrodynamical simulations \citep{Kley_Haghighipour2014} and investigated in more detail using analytical approaches and N-body simulations \citep{Zoppetti+2018,Sutherland_Kratter_2019,Gianuzzi+2023}. However, the stability of these resonances depends on parameters such as the eccentricities of the planet and the binary star. If a resonance is disrupted, planetary migration may resume or the planet may be ejected from the system. \cite{Martin_Fitzmaurice2022} shows that very low-mass planets (around one Earth mass) migrate more slowly and are therefore more likely to be captured into resonances, where they are more vulnerable to dynamical instability and ejection. In contrast, more massive planets are more likely to remain in stable resonant configurations, provided that disc turbulence is not too strong.

Furthermore, MMRs can form between two CBPs during or after their migration within the protoplanetary disc in addition to those with the binary star. In the case of Kepler-34, \cite{Kley_Haghighipour2015} showed that because of planet-planet scattering, the scenario of two planets migrating and entering together in resonance could give results more similar to observations than the scenario of only one planet migrating. Moreover, depending on the planet–planet mass ratio, the final planetary architecture could significantly change in the inner part of the protoplanetary disc, related to MMRs. In addition, \cite{Fitzmaurice+2022} suggest that the observed lack of low-mass planets, as reported by \cite{Martin_2018}, may be partially explained by MMRs between planets. These resonances can either lead to the preferential ejection of rocky planets or trap them on longer-period orbits (often in resonance with a more massive inner planet), making them significantly harder to detect.

The goal of this work is to investigate how binary star parameters influence the stability of migrating low-mass multi-planet circumbinary systems using N-body simulations. The very small number of observed systems hosting multiple low-mass circumbinary planets cannot be explained solely by observational biases \citep{Martin_Triaud2014}. In addition, recent studies based on the Exoplanet Encyclopaedia reveal a pronounced deficit of relatively close binary systems among planet-hosting stars, suggesting that binarity itself may hinder planet formation \citep{Thebault_Bonanni2025}. Despite these indications, circumbinary planet formation has rarely been explored beyond the limited parameter space of currently observed systems, and the wide diversity of binary properties remains largely untested.

In order to extend previous works such as \cite{Martin_Fitzmaurice2022}, \cite{Fitzmaurice+2022} and \cite{Gianuzzi+2023}, we investigate the dynamical stability of two-planet circumbinary systems. In particular, we explore how variations in the binary eccentricity and the binary mass ratio affect the final system architecture. We also compare the outcomes of systems in which planets become trapped in planet-planet MMR during their migration with respect to those in which planets migrate independently.

The paper is organised as follows: in Section \ref{sec:Methodology}, we present the simulation setup and the whole set of simulation parameters used; in Section \ref{sec:Results}, we present the results of the series of simulations and our analysis; in Section \ref{sec:Discussion}, we discuss the validity of our model and compare our results with previous work. We conclude and list our main findings in Section \ref{sec:Conclusions}.

\section{Methodology}
\label{sec:Methodology}

To investigate the effects of binary eccentricity and binary mass ratio on the final architecture of planets, we simulate the migration phase during which planets undergo orbital evolution within the protoplanetary disc shortly after their formation.

\subsection{Simulation set}
\label{subsec:Simulation set}

\begin{figure}
    \centering
    \includegraphics[width=0.45\textwidth]{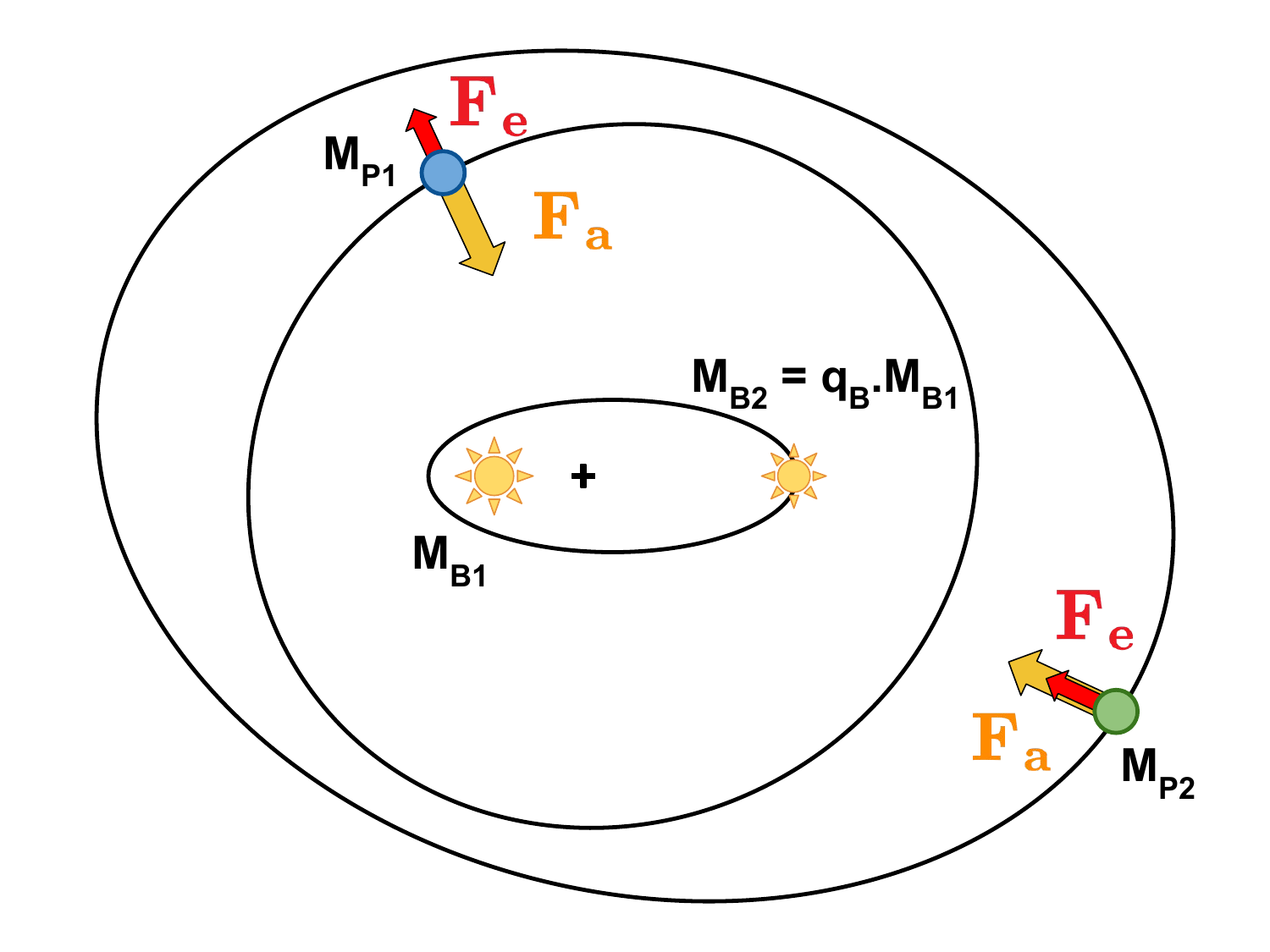}
    \caption{Schematic of the simulation setup described in Section~\ref{subsec:Simulation set}. Around a binary star with masses $M_{B1}$ and $M_{B2}$, we set two planets, the innermost one with a mass $M_{P1}$ and the outermost one with a mass $M_{P2}$. Each planet undergoes two external forces, $\vec{F_a}$ and $\vec{F_e}$, defined in Equation~\ref{eq:migration_forces}.}
    \label{fig:Simulation_setup}
\end{figure}

We perform N-body simulations by using the REBOUND C module \citep{Rein_Liu2012} and the IAS15 adaptative integrator without any limitation on time steps \citep{Rein_Spiegel2015}. Figure~\ref{fig:Simulation_setup} presents a schematic of the simulated configuration, which is described in more detail in the following paragraphs. We consider a binary star with a separation equals to $a_B$ = 0.2 AU and a total mass $M_B$ = $M_{B1} + M_{B2}$ = 1 $M_{\odot}$. We sample ten values of the binary mass ratio $q_B = {M_{B2} / M_{B1}}$ between 0.1 and 1.0 (with 0.1 steps), considering six different values for the binary eccentricity $e_B$ from 0 to 0.5 (with 0.1 steps). We place two planets with the same mass $M_{P1} = M_{P2} = 10 \ M_{\oplus}$. Initially, we set their eccentricities $e_{P1} = e_{P2} = 0$.

We aim to determine whether the timing at which the planets become close enough to undergo strong mutual interactions (either during migration or at its end) has an impact on the final outcome. To do so, we set two different series of simulations in order to study these interactions during the migration. 

In the first setup, the semi-major axis of Planet~1 $a_{P1}$ is initially set to 6 AU from the binary’s centre of mass, while Planet~2 is placed at $a_{P2}$ = 10.15 AU. In this configuration, the mean-motion ratio between the two planets is approximately 2.2, close to the 2:1 resonance. If the planets undergo convergent migration (i.e., they move toward each other during migration), these initial positions allow a stable 2:1 MMR to form early enough, before the innermost planet encounters a potential MMR with the binary. As a result, the planets migrate as a resonantly coupled pair, remaining in strong mutual interaction throughout a significant fraction of their migration. For this reason, we refer to this configuration as the "simultaneous migration" scenario.

For the second series of simulations, all parameters are kept identical, except that Planet 2 is initially placed at $a_{P2} = 30$~AU. In this configuration, the planets still undergo convergent migration. However, because their initial separation is significantly larger, resonant capture is expected to occur at a much later stage. As a result, Planet 1 reaches its parking location without being captured into resonance with Planet 2, and the two planets do not experience strong mutual interactions until the end of Planet 2’s migration. We refer to this configuration as the "sequential migration" scenario, as the planets migrate almost independently, allowing the simulation to be interpreted as two successive phases of single-planet migration leading to the formation of a potential multi-planet architecture.

We run a total of 1200 simulations: for each combination of $(q_B, e_B,\text{migration scenario})$, we perform 10 simulations with random initial orbital phases. In particular, the argument of pericentre $\omega$ and the true anomaly $\nu$ of each body are drawn from uniform distributions in $[0^{\circ},360^{\circ}]$. Each simulation is integrated over 1 Myr, ensuring that the planets have sufficient time to migrate toward the inner region, experience mutual and binary-induced interactions, and possibly establish MMRs. This duration further enables us to evaluate the long-term stability of any resonant configurations over several hundred thousand years. With our parameter values, the binary orbital period is $P_B \approx 33$ days, while planets in MMR with the binary typically have orbital periods on the order of $P_P \approx 1$ year.

\subsection{Migration model}
\label{subsec:Migration model}

To reproduce the migration of low-mass planets within a circumstellar protoplanetary disc, several prescriptions have been developed for Type-I migrations \citep[e.g.][]{Cresswell_Nelson2008,Coleman_Nelson2014,Ida+2020}. These prescriptions consist in adding external forces acting on planets, which make them migrate inward: forces depend on planet mass $M_P$, velocity $\vec{v}$, separation between the planet and the centre of mass $\vec{r}$, and characteristic times for the orbital migration $\tau_a$, for the planet's eccentricity damping $\tau_e$, and for the planet's inclination damping $\tau_i$. For a realistic prescription, $\tau_a$, $\tau_e$ and $\tau_i$ depend on disc properties as well as stellar and planetary orbital parameters. However, in order to adopt a general approach and to favour the establishment of convergent migration, we use a simplified prescription with constant characteristic timescales, chosen arbitrarily and kept fixed throughout the simulations. The expressions adopted for the external forces, following \cite{Zoppetti+2018}, are given by the following equation:
\begin{equation}
        \vec{F_a} = - \frac{M_{P}.\vec{v}}{2.\tau_{a}} \ ; \ \vec{F_e} = -2 \frac{M_{P}.(\vec{v}.\vec{r})\vec{r}}{r^{2}.\tau_e}.
\label{eq:migration_forces}        
\end{equation}
This migration prescription, based on \cite{Cresswell_Nelson2008}, works well when considering low planet eccentricity during the whole migration. However, according to \cite{Zoppetti_2019}, circumbinary planets undergo secular evolution for their eccentricity, which can lead to high eccentricity values and thus invalidate the previous assumption. To account for this, we modify the expression of $\tau_{a}$ following \cite{Goldreich_Schlichting_2014}:
\begin{equation}
    \frac{1}{\tau_{a_{modif}}} = \frac{1}{\tau_a} + \frac{2 \ \beta \ {e_P}^2}{\tau_e} ,
\label{eq:modification_Goldreich}
\end{equation}
where $e_P$ is the planet eccentricity, and $\beta$ is a
factor that quantifies the fraction of the orbital angular momentum
preserved during the migration. We choose $\beta = 0.3$ as proposed by \cite{Goldreich_Schlichting_2014} according to estimations of \cite{Tanaka_Ward_2004}. We do not implement an inclination damping force since the planets, the binary and the disc are in the same plane in our simulations. We assume that the circumbinary disc remains circular and axisymmetric, and thus do not account for any azimuthal asymmetries in the applied external forces. Whenever the eccentricity becomes higher than 1, we consider that the planet is ejected from the system and we remove the external forces caused by the escaping planet.

In our simulations, we adopt values of $\tau_a$ and $\tau_e$ consistent with the characteristic timescales expected for Type-I migration of a planet with our adopted mass, assuming typical disc parameters near the inner edge of the disc. We consider a surface density profile $\Sigma(r) = 487.74 \ \text{kg m}^{-2} . (r / 1 \text{AU})^{-1}$ and an aspect ratio profile $H(r)/r = 0.0283 \ .(r / 1 \ \text{AU} )^{0.25}$. Using the prescription for migration and eccentricity damping timescales from \cite{Cresswell_Nelson2008}, we obtain for a 10 $M_{\oplus}$, non-eccentric planet at 1 AU characteristic values of $\tau_a \approx 2 \times 10^{5}$ years and $\tau_e \approx 800$ years. We further select the innermost and outermost planet's migration timescales, $\tau_{a,i}$ and $\tau_{a,o}$, such that resonant capture between the planets occurs before either planet reaches a potential MMR with the binary. In addition, we note that the planets enter in resonance before the eccentricity forcing induced by the secular effects of the binary becomes too strong. Such forcing can significantly alter the amplitude of the migration torques and potentially prevent convergent migration, thereby hindering the establishment of a stable MMR between the two planets. Using these migration prescriptions, the semi-major axes evolves according to an exponential decay:
\begin{equation}
    a_i(t) = a_{i,0}.\exp\left(-\frac{t}{\tau_{a,i}}\right); \\
    a_o(t) = a_{o,0}.\exp\left(-\frac{t}{\tau_{a,o}}\right). 
\label{eq:exponential decreasing}
\end{equation}

Thus, we adopt $\tau_{a,i} = 2.2 \times 10^{5}$ years, $\tau_{a,o} = 1.8 \times 10^{5}$ years, and $\tau_e$ = 1,000 years for both planets. Following the analysis of circumstellar systems by \cite{Lin+2025}, an important parameter controlling the stability of planetary MMRs is the characteristic differential migration timescale $\Delta\tau_a$ defined as follows:
\begin{equation}
    {\Delta\tau_a}^{-1} = {\tau_{a,o}}^{-1} - {\tau_{a,i}}^{-1}.
\label{eq:Delta_taua}
\end{equation}
For our adopted values, we obtain $\Delta\tau_a \approx 10^{6}$ years and $\Delta\tau_a / \tau_e \approx 1,000$. According to the criteria of \cite{Lin+2025}, these values should generate a long-term MMR between planets. The adopted values of $\tau_{a,i}$, $\tau_{a,o}$ and $\tau_e$ are also consistent with the timescales computed with the prescription from \cite{Cresswell_Nelson2008}. Finally, we control the expected mean eccentricity of the inner planet at the time of resonant capture between the two planets. Using the expressions from Eq.~\ref{eq:exponential decreasing} and the adopted characteristic timescales, resonant capture is expected to occur at approximately 4.5 AU, well before any potential N:1 resonances with the binary star. For all considered pairs of ($q_B$,$e_B$), we computed the capture mean eccentricity of the inner planet at 4.5 AU (using Eq.~27 in \citealt{Zoppetti_2019}) and verified that it does not exceed 0.02. The resulting variation in the amplitudes of the forces $\vec{F_a}$ and $\vec{F_e}$ relative to the circular case therefore remains below 5\%, which we consider satisfactory.

\section{Results}
\label{sec:Results}

\subsection{Overview of the orbital evolution of circumbinary planets migrating}
\label{subsec:overview_orbital_evolution}

\begin{figure*}[t]
\centering

\begin{tabular}{cc}
\begin{overpic}[width=0.4\textwidth]{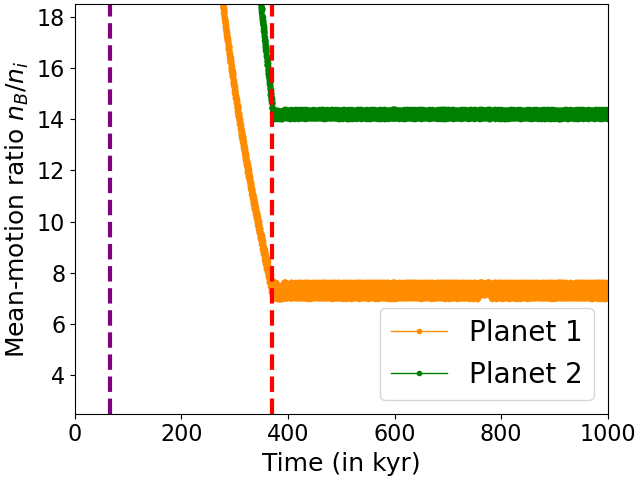}
\put(80,52){\bfseries (a)}
\end{overpic}
&
\begin{overpic}[width=0.4\textwidth]{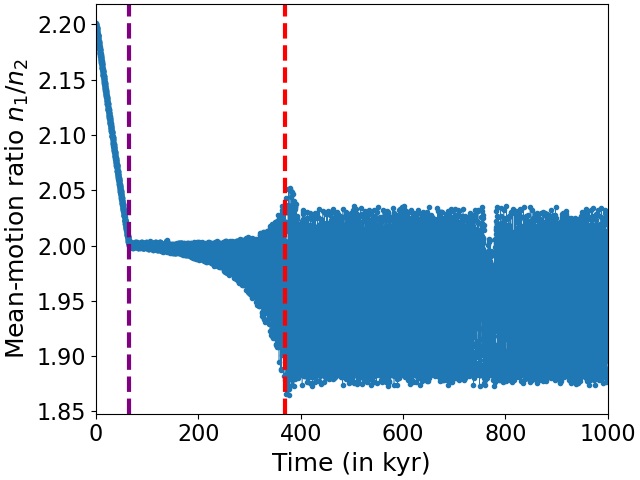}
\put(80,52){\bfseries (b)}
\end{overpic}
\\

\begin{overpic}[width=0.4\textwidth]{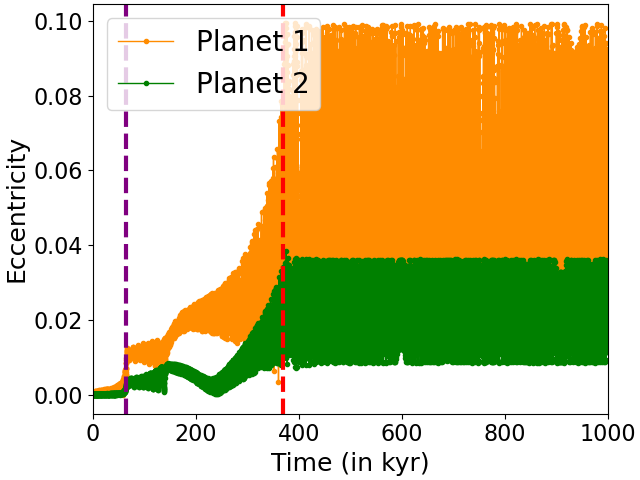}
\put(80,52){\bfseries (c)}
\end{overpic}
&
\begin{overpic}[width=0.4\textwidth]{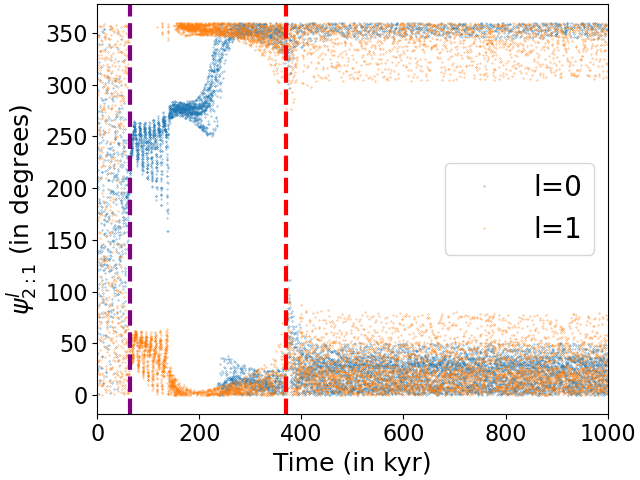}
\put(80,52){\bfseries (d)}
\end{overpic}
\end{tabular}

\caption{Evolution of several physical quantities in a simulation with $q_B=0.6$, $e_B = 0.5$ in the framework of the simultaneous migration scenario. Panel~(a): mean-motion ratios between the planets and the binary star. Panel~(b): mean-motion ratio between both planets. Panel~(c): eccentricity of both planets. Panel~(d): resonant angles related to a 2:1 resonance between planets. The purple vertical dashed line represents the moment when both planets enter MMR, and the red vertical dashed line represents the time when Planet 1 enters a 7:1 resonance with the binary.}
\label{fig:Sim_qB0.6_eB0.5_simult}

\end{figure*}

\begin{figure*}[t]
\centering

\begin{tabular}{cc}
\begin{overpic}[width=0.4\textwidth]{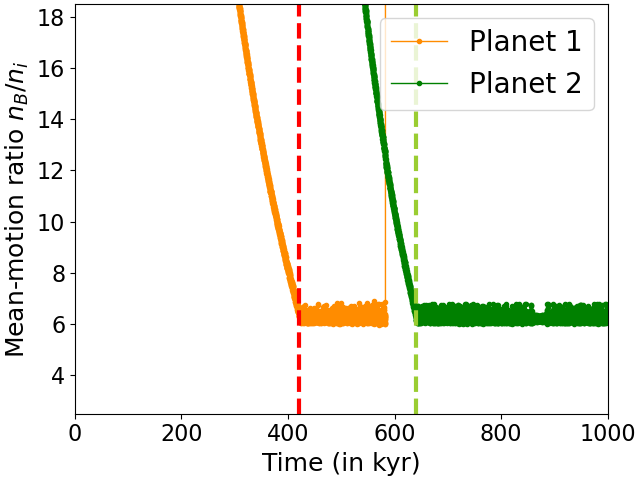}
\put(80,52){\bfseries (a)}
\end{overpic}
&
\begin{overpic}[width=0.4\textwidth]{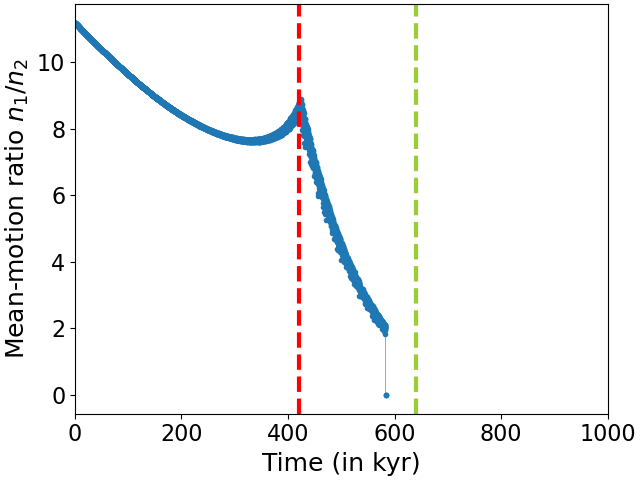}
\put(80,52){\bfseries (b)}
\end{overpic}
\\

\begin{overpic}[width=0.4\textwidth]{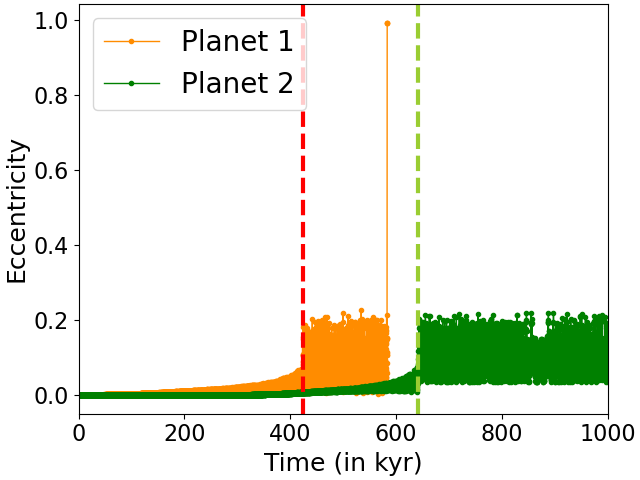}
\put(80,52){\bfseries (c)}
\end{overpic}
&
\begin{overpic}[width=0.4\textwidth]{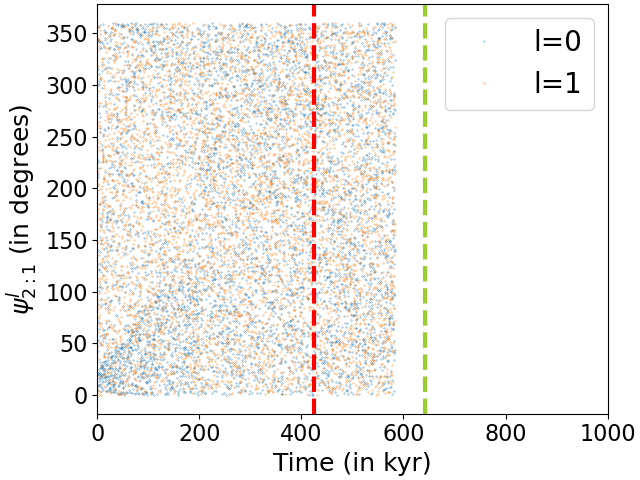}
\put(80,52){\bfseries (d)}
\end{overpic}
\end{tabular}

\caption{Same as Figure \ref{fig:Sim_qB0.6_eB0.5_simult}, considering $q_B=0.5$, $e_B = 0.3$ in the framework of the sequential migration scenario. The red vertical dashed line represents the moment when Planet~1 enters a 6:1 resonance with the binary while the green vertical dashed line represents the time when Planet~2 enters a 6:1 resonance with the binary.}
\label{fig:Sim_qB0.5_eB0.3_sequent}

\end{figure*}

\begin{figure*}[t]
\centering

\begin{tabular}{cc}
\begin{overpic}[width=0.4\textwidth]{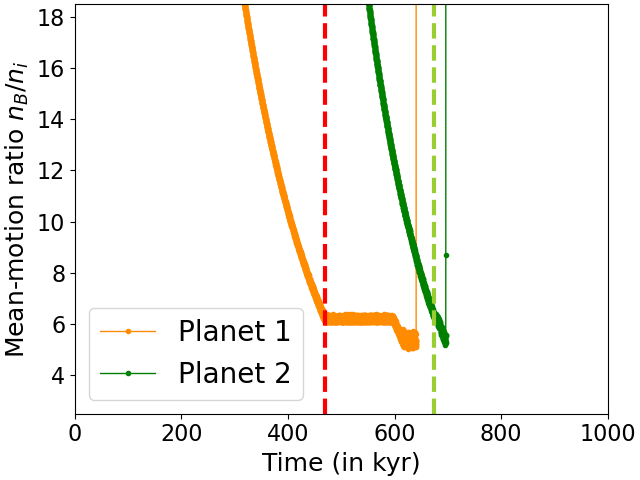}
\put(80,52){\bfseries (a)}
\end{overpic}
&
\begin{overpic}[width=0.4\textwidth]{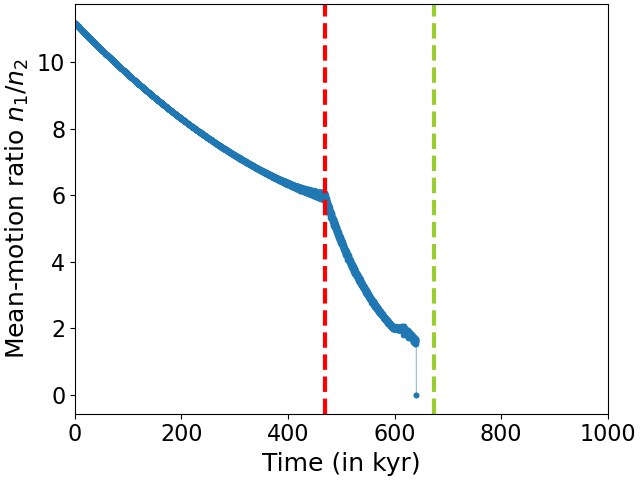}
\put(80,52){\bfseries (b)}
\end{overpic}
\\

\begin{overpic}[width=0.4\textwidth]{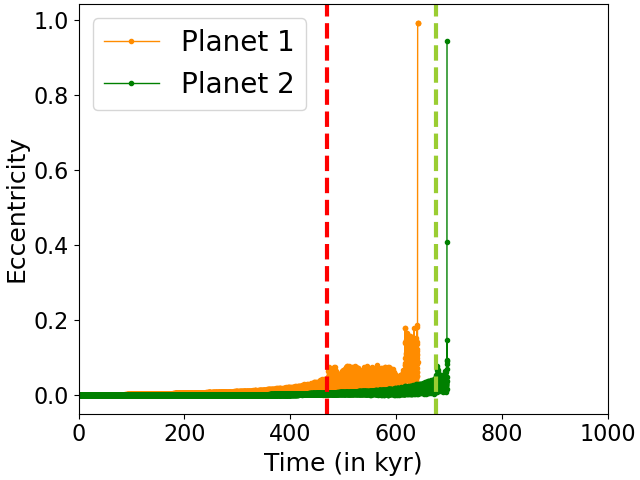}
\put(80,52){\bfseries (c)}
\end{overpic}
&
\begin{overpic}[width=0.4\textwidth]{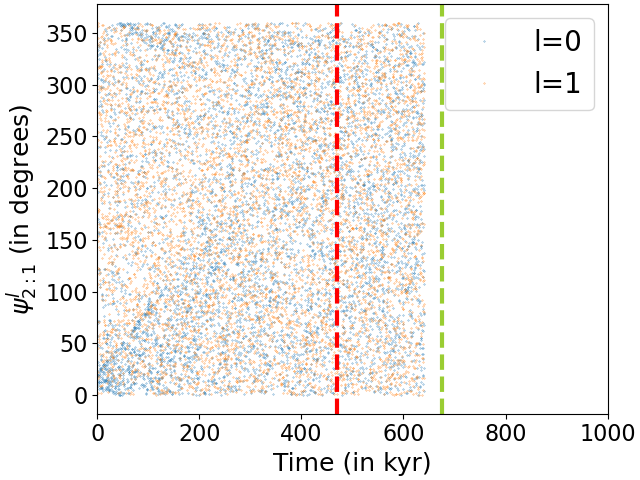}
\put(80,52){\bfseries (d)}
\end{overpic}
\end{tabular}

\caption{Same as Figures \ref{fig:Sim_qB0.6_eB0.5_simult} and \ref{fig:Sim_qB0.5_eB0.3_sequent}, considering $q_B=0.7$, $e_B = 0.2$ in the framework of the sequential migration scenario. The red vertical dashed line represents the moment when Planet~1 stops its migration at $n_B/n_1$ = 6.2, and the green vertical dashed line represents the time when Planet~2 stops its migration at $n_B/n_2$ = 6.2.}
\label{fig:Sim_qB0.7_eB0.2_sequent}

\end{figure*}

In all simulations, we observe a similar sequence of evolutionary phases: an inward migration of both planets, followed by their capture into MMR, either with each other or with the binary star. In the majority of simulations, the planets do not remain stable until the end of the integration. Instead, they often undergo close encounters with the binary or with the companion planet, which almost invariably leads to their ejection from the inner circumbinary region.

Figure~\ref{fig:Sim_qB0.6_eB0.5_simult} shows the temporal evolution of the most relevant physical quantities for a representative simulation involving two interacting circumbinary planets migrating around a binary star with $q_B = 0.6$ and $e_B = 0.5$. Panel~(a) displays the mean-motion ratios between the binary and the planets, defined as $n_B/n_i$, where $n_B$ and $n_i$ denote the mean motions of the binary and Planet~i. They illustrate the inward migration of both planets and their possible capture into MMRs with the binary. Panel~(b) shows the mean-motion ratio between the two planets $n_1/n_2$, indicating whether they become locked in mutual MMR during or after migration. Panel~(c) presents the eccentricity evolution of both planets, while Panel~(d) shows the resonant angles associated with the 2:1 planet-planet MMR.

The resonant angles with respect to a \textit{p:q} MMR between two bodies (\textit{p}, \textit{q}, two small integers) are:
\begin{equation}
    \psi_{p:q}^{l} = p.\lambda_{a} - q.\lambda_{b} - (p-q).\varpi_{a} + l.(\varpi_{a} - \varpi_{b}) \,\,\,,
    \label{eq:resonant_angles}
\end{equation}
where $\lambda_{a}$ (resp. $\lambda_{b}$) is the mean longitude of the outer body (resp. the inner body), $\varpi_{a}$ (resp. $\varpi_{b}$) is the longitude of pericentre of the outer body (resp. the inner body) and $l$ an integer between 0 and $p-q+1$. 
Resonant angles are the only reliable way to confirm the presence of a MMR. While a ratio $n_a/n_b \approx p/q$ may suggest a possible commensurability, it does not by itself imply resonance. During a MMR, at least one resonant angle is expected to converge toward a constant value and then librate around it. Thus, a system can only be considered to be in a true \textit{p:q} MMR if at least one of the associated resonant angles exhibits such libration.

At the beginning of the simulation illustrated in Figure~\ref{fig:Sim_qB0.6_eB0.5_simult}, planetary migration starts immediately, as evidenced by the decrease in the planets’ semi-major axes (and thus mean-motion ratios with the binary as observed in Panel~(a)). Oscillations of all parameters of Figure~\ref{fig:Sim_qB0.6_eB0.5_simult} are typically due to perturbations from the binary motion, in addition to planet–planet interactions to a lesser extent.

Panel~(b) also shows the convergent migration of both planets related to different migration timescales: $n_1/n_2$ decreases down to 2. Starting from $t = 65$ kyr, it oscillates around 2, suggesting both planets enter in a 2:1 MMR. Panel~(d) confirms this behaviour. At early times ($t \lesssim 60-70$ kyr), the 2:1 MMR resonant angles clearly circulate, as indicated by the dense cloud of points uniformly spanning the full 0–360° range. After this initial phase, at $t = 70$ kyr, the distribution progressively restructures and each angle begins to converge toward one preferred value (modulo 360°). This marks the transition from full circulation to libration. Another feature associated with the 2:1 MMR between the planets is a small eccentricity jump. Indeed, we observe at $t = 65$ kyr a sharp increase of approximately 0.008 for Planet~1 and 0.003 for Planet~2. The eccentricities then oscillate around nearly constant mean values until $t = 140$ kyr, when their evolution diverges. While the eccentricity of Planet~1 progressively increases, with a growing oscillation amplitude, Planet~2 first decreases to about 0.002 and then rises again around $t = 140$ kyr, also exhibiting oscillations of increasing amplitude. We suggest that the main long-term increase in the eccentricities of both planets is driven by secular effects, as described in \cite{Zoppetti_2019}. We discuss these secular effects in more details in Section~\ref{subsec:Explaining_dynamics_resonant_instabilities}.

Around $t = 370$ kyr, the innermost planet is already located in the region close to the binary, as evidenced by the large amplitude eccentricity oscillations in Panel~(c), and enter in a 7:1 MMR with the binary, as suggested in Panel~(a) with the abrupt halt in the decrease of $n_B/n_1$ around 7.25. Even though the mean-motion ratio does not oscillate perfectly around 7, the verification of resonant angles related to a 7:1 resonance enables us to observe that at least three resonant angle librate around a constant value, and thus to confirm the observed MMR (see Panel~(a) of Fig.~\ref{app:angles_combined}). The outermost planet also stops its migration shortly after, and planets remain bounded and locked in their initial 2:1 resonance. Higher amplitudes of $n_1/n_2$ can be observed in Panel~(b), but the fact that the resonant angles still librate around a fixed value confirms the continuity of the resonance. Until the end of the simulation, all parameters oscillate around the same fixed value, indicating the system stability over several hundred thousand years.

The simulation described in Figure~\ref{fig:Sim_qB0.6_eB0.5_simult} represents a case which forms a stable multi-planet system. But this case does not reflect the behaviour seen in most of the simulations performed. Indeed, 286 out of 1200 simulations lead to two surviving planets. Figure~\ref{fig:Sim_qB0.5_eB0.3_sequent} shows the most common case in both migration scenarios (with 756 out of 1200 simulations). In this simulation, planets migrate independently, as evidenced by the large, non-constant value of $n_1/n_2$ seen in Panel~(b). Around $t = 420$ kyr, the innermost planet enters in a 6:1 MMR with the binary and remains stable for less than 200 kyr, as shown in Panel~(a). We also observe that from $t = 320$ kyr to 420 kyr, $n_1/n_2$ increases instead of having a convergent migration. This increase in the mean-motion ratio appears to coincide with a growth in the eccentricity of Planet~1, which may be driven by secular perturbations induced by the central binary. Because of the migration prescription, increasing planet eccentricity reduces the $\tau_{a_{modif}}$ value, which makes the planet migrate faster. Even though $\tau_{a,o} < \tau_{a,i}$, the higher eccentricity of Planet~1 may lead to a faster migration than Planet~2, explaining this increase in $n_1/n_2$. After the 6:1 resonance, we recover the expected behaviour of a convergent migration, which is more pronounced because the innermost planet is parked. At $t = 580$ kyr, $n_1/n_2$ remains constant around 2 very briefly, but we cannot assure it is a true MMR, since we do not observe any libration of resonant angles during this period. Then, the innermost planet does not remain bound to the binary and is ejected. This corresponds to an eccentricity value higher than 1. In our simulations, we do not consider any collision: therefore, we do not make the distinction between an ejection and a collision with another celestial body. The outermost planet continues its migration until it arrives at the same location as the innermost planet before the perturbation at $t = 640$ kyr. Then, the planet remains stable until the end of the simulation.

Last, 158 out of 1200 simulations lead to the ejection of both planets. Figure~\ref{fig:Sim_qB0.7_eB0.2_sequent} shows an example of this scenario: in Panel~(a), we can see both planets migrate independently. Planet~1 stops its migration when $n_B/n_1$ reaches a value around 6.2, but we notice there is not any resonant angle which librates around a constant value. In other words, Planet~1 stops its migration without entering into a true 6:1 MMR. Then, the planet remains stable until $n_1/n_2$ reaches a value around 2 at $t = 595$ kyr. Then, we notice that Planet~1 migrates again (probably because the second planet gets much closer and push the first one inwards) until $n_B/n_1$ reaches a value around 5.3 at $t = 615$ kyr. Even though there is no evidence for libration of resonant angles as shown in Panel~(d), $n_1/n_2$ remains around 2 until $t = 615$ kyr where Planet~2 continues to migrate. When $n_1/n_2$ reaches a value of 1.66 (close to a 5:3 MMR between planets even if no evidence for libration of resonant angles), Planet~1 is quickly ejected from the system. Contrary to the case in Fig.~\ref{fig:Sim_qB0.5_eB0.3_sequent}, Planet~2 stops its migration at the very first Planet~1's parking place (the 6:1 MMR with the binary), but does not remain stable more than 5 kyr and continue to migrate until $n_B/n_2$ reaches a value close to 5.3 at $t = 695$ kyr. In Panel~(c), at this period, we observe a very high increase in eccentricity up to 0.94 just before the ejection of Planet~2. This evolution likely reflects a difference in dynamical behaviour between the two planets. Planet~1 remains stable close to a 6:1 MMR, and interactions with Planet~2 are likely required to disrupt this stability. In contrast, Planet~2 may be ejected independently, without strong dynamical interactions with Planet~1.
Two possible explanations may account for this behaviour. First, differences in the migration timescales of the two planets could result in different sensitivities to MMRs \citep[see][]{Sutherland_Kratter_2019}. Alternatively, prior planet–planet interactions may have induced an additional instability in Planet~2 (e.g., a higher eccentricity), making it more prone to ejection than an isolated planet. We favour the first explanation, as we do not observe any significant increase in the eccentricity of Planet~2 at the time the stability of Planet~1 is disrupted or when Planet~1 is ejected.

Finally, we observe a small number of simulations in which an orbital swap occurs between the two planets. By orbital swap, we refer to a configuration in which the planets cross their orbits and exchange their radial ordering during migration, such that the initially outer planet becomes the inner one. Across the full set of simulations, such events remain rare, with only 9 cases in the sequential scenario and 12 cases in the simultaneous scenario. Nevertheless, all orbital swap configurations systematically lead to instability, typically resulting in the ejection of Planet~2 and the survival of Planet~1. This appears to be the only pathway in which Planet~1 remains the sole bound planet. In all other cases, either both planets survive or only Planet~2 survives.

\subsection{Impact of binary parameters on stability of circumbinary planets}
\label{subsec:impact_binary_parameters_stability}

\begin{figure*}[t]
\centering
    \begin{tabular}{cc}
        \begin{overpic}[width=0.5\textwidth]{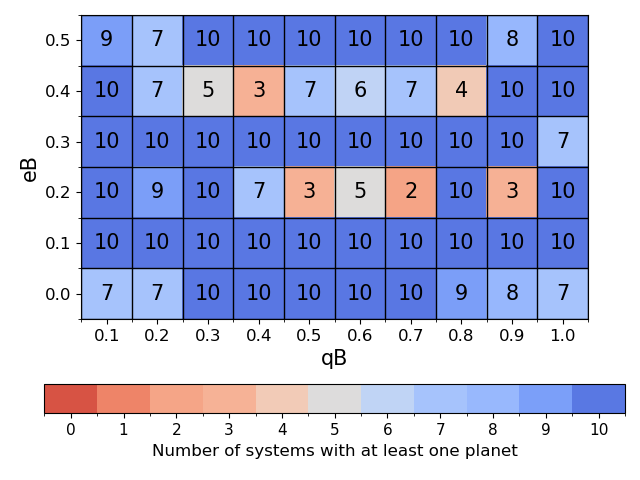}
        \put(50,-1){\bfseries (a)}
        \end{overpic} &
        \begin{overpic}[width=0.5\textwidth]{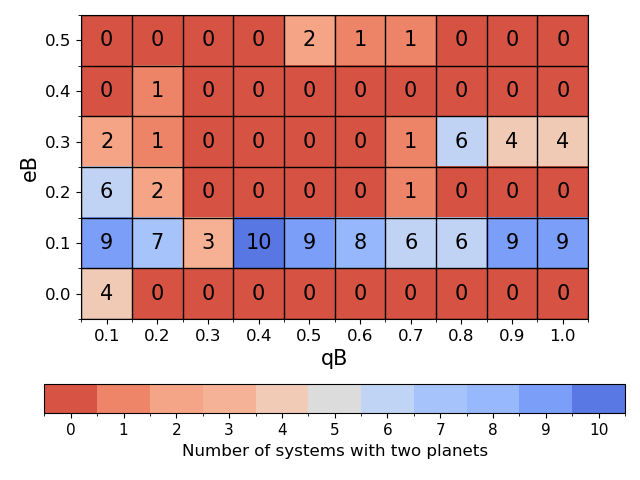}
        \put(50,-1){\bfseries (b)}
        \end{overpic}
    \end{tabular}
    \caption{
        Panel~(a): Stability maps investigating the stability of systems keeping at least one of both planets bound to the binary star. Panel~(b): Stability map investigating the stability of systems keeping both planets stable. The x-axis (resp. the y-axis) indicates the binary mass ratio (resp. binary eccentricity) studied. For each $q_B-e_B$ configuration, we consider the ten different simulations run with randomly chosen orbital angles for the planets. All the simulations consider two 10 $M_{\oplus}$ circumbinary planets that migrate following the sequential migration scenario.
    }
    \label{fig:stabilitymap1}
    
\end{figure*}

\begin{figure*}
\centering
    \begin{tabular}{cc}
        \begin{overpic}[width=0.5\textwidth]{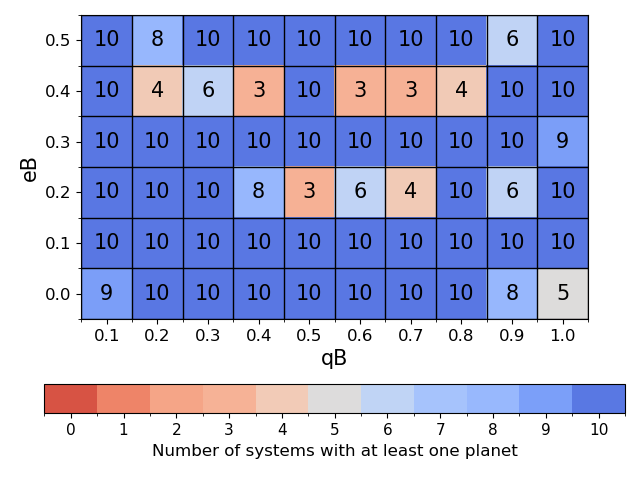}
        \put(50,-1){\bfseries (a)}
        \end{overpic} &
        \begin{overpic}[width=0.5\textwidth]{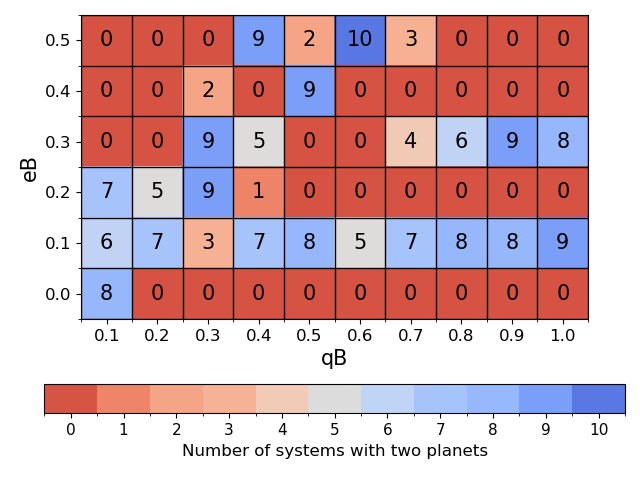}
        \put(50,-1){\bfseries (b)}
        \end{overpic}
    \end{tabular}
    \caption{
        Same as Figure \ref{fig:stabilitymap1}, considering the simultaneous migration scenario.
    }
    \label{fig:stabilitymap2}
    
\end{figure*}

To show the influence of the binary mass ratio $q_B$ and the binary eccentricity $e_B$ on the outcomes of the simulations, we made the following analysis. For each combination of parameters ($q_B$, $e_B$, migration scenario), we consider the outcomes of the ten simulations performed with different random initial phases and for each of them, we observe the final value of planets eccentricity to check which one have been ejected during the simulation (i.e. planets with eccentricity higher than 1). Then, this allows us to count the number of simulations which leads to two possible outcomes: 1) at least one planet remains stable in the system at the end of the simulation; 2) both planets remain stable. These numbers are listed in stability maps as a function of $q_B$ and $e_B$. Figure~\ref{fig:stabilitymap1} shows stability maps associated with the sequential migration scenario and Figure~\ref{fig:stabilitymap2} shows stability maps associated with the simultaneous migration scenario. A colour is assigned to each cell of the stability maps, depending on the number of simulations in which at least one planet (or both planets) remains stable until the end of the simulation. Assuming a uniform distribution of the bodies’ orbital phases, the colour represents the likelihood that a given combination of ($q_B$,$e_B$, migration scenario) produces stable architectures with one or two planets.

Regarding the sequential migration scenario, Panel~(a) of Fig.~\ref{fig:stabilitymap1} presents the stability map for systems in which at least one planet survives. The distribution is non-uniform, yet the systems appear broadly stable across the entire parameter space. In most cases, more than seven out of ten simulations result in at least one surviving planet. Some regions of reduced stability can nevertheless be identified, notably around $e_B = 0.2$ and $e_B = 0.4$, where certain ($q_B$, $e_B$) combinations yield fewer than 5 surviving systems out of 10. These outcomes show a strong dependence on the initial orbital phases of the bodies. To obtain more robust statistical results, a larger number of simulations with randomly selected initial phases would be required. This is however beyond the scope of the present work.

In Panel~(b) of Fig.~\ref{fig:stabilitymap1}, the stability map for systems in which both planets survive shows fewer stable configurations than in the previous panel. The binaries with $e_B$ = 0.1 are the only ones which give more than 7 simulations out of 10 with both planets surviving. Additional regions of reduced stability (with between 4 and 6 surviving cases out of ten) also emerge. These are first observed for binaries with a low mass ratio and relatively low eccentricity, but also for systems with a higher binary mass ratio at $e_B = 0.3$. We remain cautious in interpreting these values, as they may vary significantly depending on the initial orbital phases, but also on the migration timescales (see Section~\ref{subsec:migration_timescales}). Overall, in the sequential migration scenario, only 112 out of 600 simulations lead to the stable formation of a multi-planet system, with 76 of these occurring primarily at $e_B = 0.1$. This result underscores the significant challenges associated with the formation and long-term stability of multi-planet systems in the context of circumbinary planet migration.

Looking at the simultaneous migration scenario, Panel~(a) of Fig.~\ref{fig:stabilitymap2} reveals stability and instability regions that are broadly similar to those identified in the sequential migration case. This similarity is further confirmed by Figure~\ref{fig:stabilitymap_diff}, which presents the residual map obtained from the difference between the stability maps shown in Figures~\ref{fig:stabilitymap1} and \ref{fig:stabilitymap2}.
In terms of statistics, 525 simulations result in at least one surviving planet in the simultaneous migration scenario, compared to 517 in the sequential case. These numbers indicate that, overall, there is no significant difference between the two migration scenarios with respect to the survival of at least one planet.

However, larger differences arise when considering only systems in which both planets remain stable. Panel~(b) of Fig.~\ref{fig:stabilitymap2} highlights configurations that are significantly more stable than those obtained in the sequential migration scenario (as also illustrated in Figure~\ref{fig:stabilitymap_diff}). The stability regions identified in the sequential case remain visible; however, an additional region of stability emerges along a diagonal extending from $q_B = 0.3$ and $e_B = 0.2$ to $q_B = 0.5$ and $e_B = 0.5$. While this feature is faintly discernible at $e_B = 0.5$ in Panel~(b) of Fig.~\ref{fig:stabilitymap1} for the sequential scenario, it becomes much more prominent in the simultaneous migration case. Overall, Panel~(b) of Fig.~\ref{fig:stabilitymap2} exhibits a higher number of stable multi-planet systems than the sequential scenario (174 compared to 112), and, more importantly, across a broader range of binary eccentricities.

\subsection{MMRs observed between planets, and between a planet and the binary star}
\label{subsec:MMR_observed_BP_PP}

\begin{figure*}
\centering
    \begin{tabular}{cc}
        \begin{overpic}[width=0.5\textwidth]{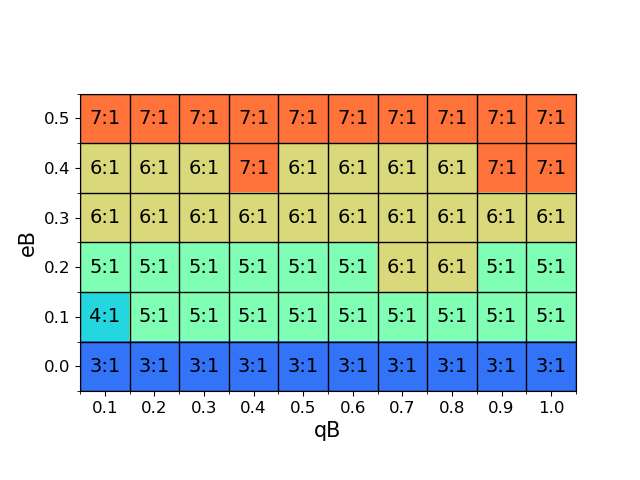}
        \put(49,-1){\bfseries (a)}
        \end{overpic} &
        \begin{overpic}[width=0.5\textwidth]{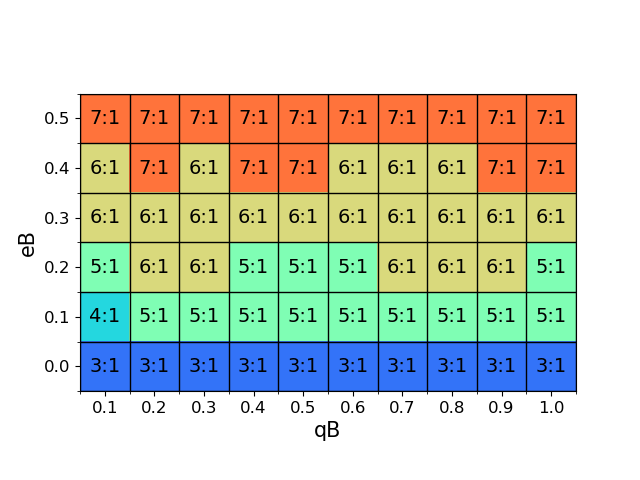}
        \put(49,-1){\bfseries (b)}
        \end{overpic}
    \end{tabular}
    \caption{
        MMR maps indicating the most frequently occuring resonance between the binary and the inner stable planet. Panel~(a): MMR map for systems in the sequential scenario. Panel~(b): MMR map for systems in the simultaneous scenario. The x-axis (resp. the y axis) indicates the binary mass ratio (resp. binary eccentricity) considered.
    }
    \label{fig:MMRmap1}
    
\end{figure*}

\begin{figure*}
\centering
    \begin{tabular}{cc}
        \begin{overpic}[width=0.5\textwidth]{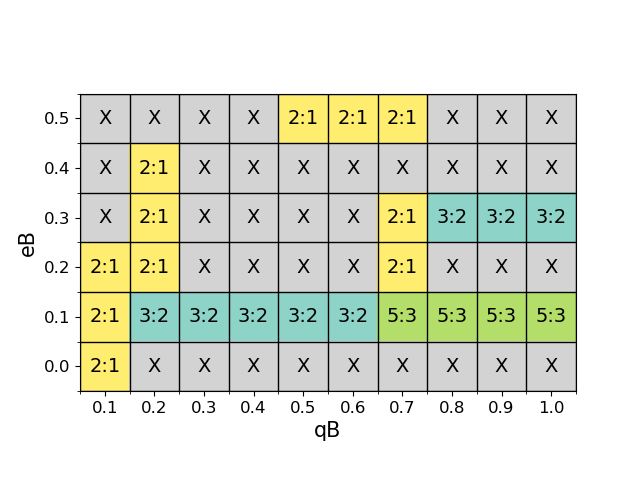}
        \put(49,-1){\bfseries (a)}
        \end{overpic} &
        \begin{overpic}[width=0.5\textwidth]{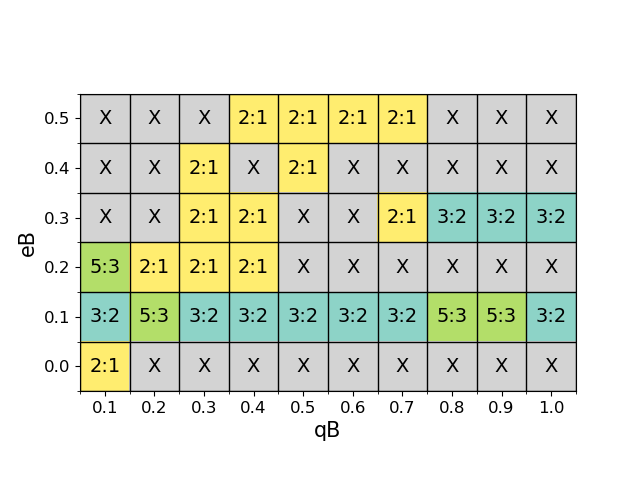}
        \put(49,-1){\bfseries (b)}
        \end{overpic}
    \end{tabular}
    \caption{
        MMR maps showing the most frequently occurring resonance between the two surviving planets. Panel~(a): MMR map for systems in the sequential scenario. Panel~(b): MMR map for systems in the simultaneous scenario. The x-axis (respectively, y-axis) represents the binary mass ratio (resp. binary eccentricity) considered. Cells marked with an “X” correspond to ($q_B$, $e_B$) configurations for which no simulation results in a stable MMR between the two planets.
    }
    \label{fig:MMRmap2}
    
\end{figure*}

In order to characterize the stable resonant chains that may form during the different simulations, we construct maps shown in Figure~\ref{fig:MMRmap1}. These maps display the most frequently occurring resonance, based on resonant angles, for the sequential migration scenario in Panel~(a), and for the simultaneous migration scenario in Panel~(b). Constructing these MMR maps is a challenging task, as the identification of a resonance cannot be reliably based solely on the values of $n_B/n_1$ and $n_B/n_2$. Instead, a robust confirmation requires the analysis of resonant angle libration. While such a verification can be performed by visual inspection for an individual simulation, this approach becomes impractical when applied to a dataset of 1,200 simulations, each of which must be tested against multiple possible resonances. We therefore developed a procedure to systematically explore all simulations and candidate resonances. This procedure is described in Appendix~\ref{app:complex_number_zl}.

In Figure~\ref{fig:MMRmap1}, each cell shows the most frequently observed resonance among the 10 associated simulations. In both panels, we observe that the resulting MMR ranges from 3:1 to 7:1 and depends primarily on the binary eccentricity: as $e_B$ increases, the value of $N$ in the N:1 MMR also increases. This trend has been reported in several studies \citep[e.g.][]{Nelson_2003,Martin_Fitzmaurice2022,Gianuzzi+2023}. In the latter, the authors found that $e_B$ is the parameter most strongly correlated with the final mean-motion ratio reached by the planet in single-planet simulations, although the MMR also depends on the binary mass ratio. However, the structure of the MMR map is similar between both migration scenarios. One can observe some local variations at $e_B$ = 0.2 and $e_B$ = 0.4 for instances. Finally, it is worth recalling that we do not include an inner disc edge in our simulations. Since planetary migration is expected to halt at the inner edge of the circumbinary disc, the formation of MMRs with the binary below this location is highly unlikely. According to \cite{Miranda_Lai_2015}, the inner edge is typically located between 2 and 3 $a_B$ for a coplanar circumbinary disc. Consequently, the 3:1 MMRs, and possibly the 4:1 MMRs, are unlikely to form in circumbinary systems, and especially the cases with $e_B$ = 0 should be interpreted with caution.

In Figure~\ref{fig:MMRmap2}, we present similar maps for the MMRs between the two planets. We follow the same methodology as the one presented in Appendix~\ref{app:complex_number_zl}, but we restrict the analysis to lower-order resonances and consider only systems with both planets remaining stable until the end of the simulations. In Panel~(a), corresponding to the sequential scenario, most ($q_B$, $e_B$) combinations lead to no MMR between the two planets across the 10 associated simulations. These cases are indicated by grey cells marked with an “X”. It is worth noting that the regions without detected MMRs in Panel~(a) do not exactly coincide with the fully unstable regions (zero simulations with two surviving planets) shown in Panel~(a) of Fig.~\ref{fig:stabilitymap1}. For instance, the case $q_B$ = 0.1 and $e_B$ = 0.3 shows no detected MMR, although two simulations retain both planets. This highlights a phenomenon that remains difficult to explain: some systems exhibit stable architectures with orbital period commensurabilities, yet without libration of the resonant angles (15 simulations in the sequential scenario and 7 in the simultaneous scenario). Focusing on the sequential scenario, and considering the ($q_B$, $e_B$) parameter space, the 2:1 MMR is the most frequent resonance, appearing as dominant in 11 cells, followed by the 3:2 MMR in 8 cells and the 5:3 MMR in 4 cells. However, when counting individual simulations instead of cells, the 3:2 MMR becomes the most common outcome, occurring in 43 simulations, compared to 33 for the 2:1 MMR and 21 for the 5:3 MMR. Panel~(b), corresponding to the simultaneous scenario, exhibits a broadly similar structure, except for the emergence of the new stable diagonal region dominated by the 2:1 MMR. In terms of individual simulations, the 2:1 resonance clearly dominates, with 85 occurrences, whereas the 3:2 and 5:3 MMRs appear in 54 and 28 simulations, respectively. This predominance of the 2:1 resonance may be explained by the migration timescales adopted from \cite{Lin+2025}, which were specifically chosen to favour the formation of a 2:1 resonance during the migration.

\subsection{Impact of migration timescales on the overall stability}
\label{subsec:migration_timescales}

\begin{figure*}
\centering
    \includegraphics[width=1\linewidth]{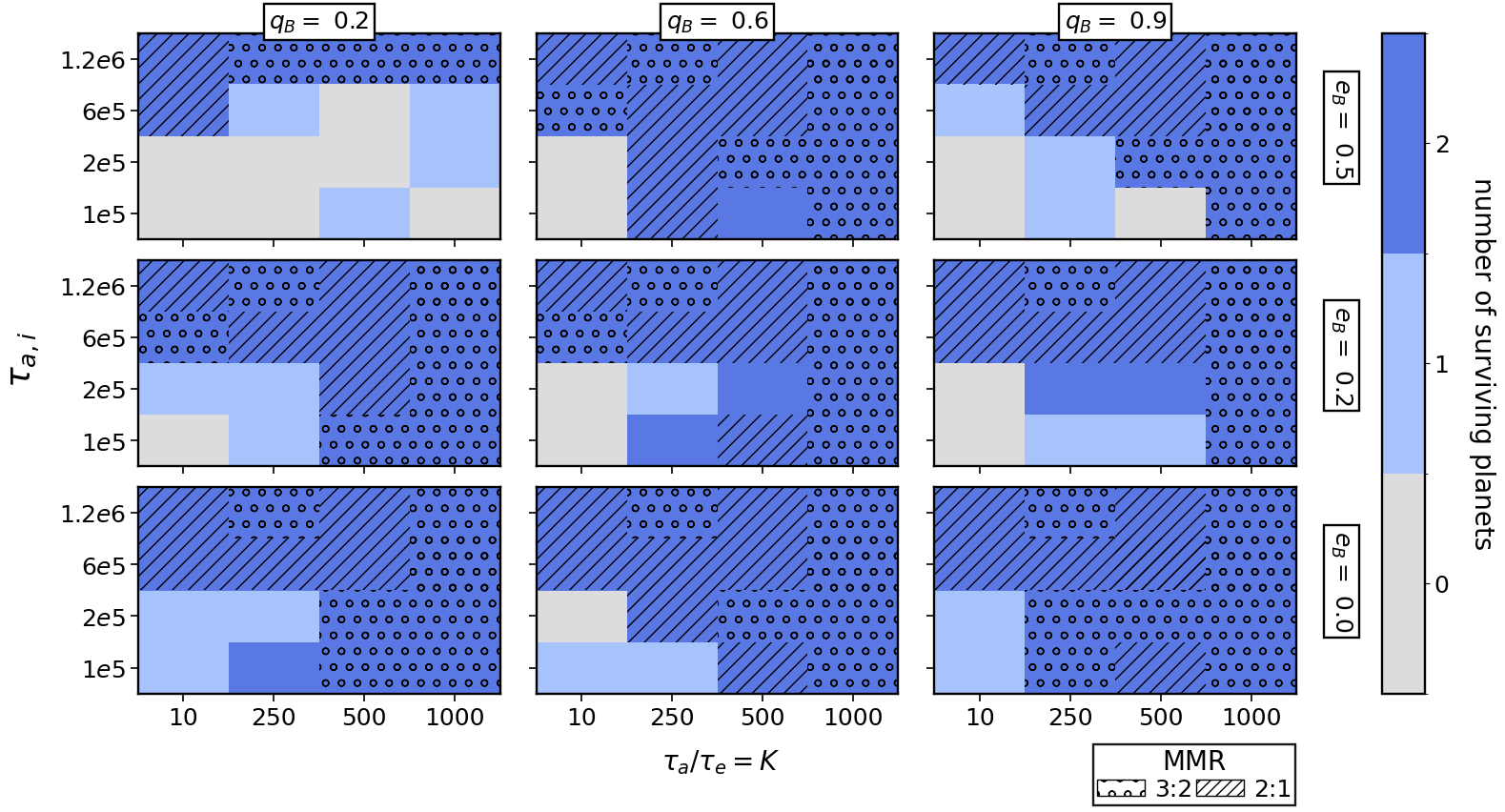}
    \caption{
    Number of planets remaining bound to the system as a function of the inner planet migration timescale $\tau_{a,i}$ (y-axis) and the ratio of semimajor axis to eccentricity damping timescales $K = \tau_a/\tau_e$ (x-axis). Shaded regions depict capture into 3:2 (dotted) or 2:1 (striped) MMRs. Panels correspond to $e_B=0.5, 0.2, 0.0$ from top to bottom, and $q_B=0.2, 0.6, 0.9$ from left to right.
    }
    \label{fig:impact_timescales}
\end{figure*}

In the previous sections, we adopted a fixed set of radial and eccentricity damping timescales ($\tau_{a,i} = 2.2 \times 10^{5}$ years, $\tau_{a,o} = 1.8 \times 10^{5}$ years, and $\tau_e$ = 1,000 years), corresponding to a differential migration timescale $\Delta \tau_{a}\approx 10^6$~years. In order to assess how robust our results are with respect to this choice, we now explore a broader range of migration regimes while keeping $\Delta \tau_a$ approximately constant. This is achieved by varying the migration timescale of the inner planets $\tau_{a,i}$ along with the ratio $K = \tau_a/\tau_e$, so that the differential migration timescale ($\Delta \tau_{a}$) between the planets is preserved, but the migration and circularization rates are modified individually. 

To this end, we consider a set of extreme binary configurations—spanning low, intermediate, and high eccentricities and stellar mass ratios—to assess the impact of migration and eccentricity damping timescales on the system. Figure~\ref{fig:impact_timescales} shows the number of surviving planets as a function of $K$ and $\tau_{a,i}$, for different binary configurations. In all cases, the planetary configuration follows the simultaneous migration scenario, which ensures that both planets migrate as a coupled pair.

A first trend is that as the eccentricity of the binary system increases, a slight decrease in the system's overall stability is observed. However, this loss of stability is not uniform; rather, it appears to be confined to a more limited region of parameter space, where the value of $q_B$ seems to influence stability, regardless of the exact values of $\tau_{a}$ and $K$. In the case of circular binary systems ($e_B = 0$), stability remains high across the entire range of $q_B$ values, and a wide variety of migration parameters leads to configurations in which both planets survive. This behaviour is explained by the fact that, since there is no eccentricity in the binary system, the perturbations experienced by the planets are weaker and more regular, which limits the excitation of eccentricity and prevents the system from reaching chaotic regions associated with the overlap of resonances.

Conversely, as $e_B$ increases, the influence of $q_B$ becomes more pronounced. In particular, for $e_B = 0.5$, configurations with strongly unequal stellar mass ratios (i.e., low and high $q_B$) exhibit markedly reduced stability compared to cases of intermediate eccentricity, as evidenced by the reduction in regions with two surviving planets and the increase in the presence of systems with only one or no survivors. In these regimes, the secular forcing exerted by the binary's eccentricity gradually increases the planet's eccentricity. As a result, the planets are more easily pushed toward regions of N:1 resonance overlap, which eventually leads to a decrease in the number of remaining CBPs.

However, the timescales of migration also influence stability, although their effect is still modulated by the parameters of the binary system. As shown in Fig.~\ref{fig:impact_timescales}, in the case of relatively rapid migration (low values of $\tau_{a,i}$), the planets reach the inner regions faster, where they are subjected to more intense perturbations from the binary system. When this is combined with low values of $K$—that is, inefficient circularization compared to radial migration—the planets tend to develop higher eccentricities, favouring configurations with a single surviving planet or even totally unstable, especially for high values of $e_B$.

As $\tau_{a,i}$ increases, migration slows down and the $e$-damping becomes more effective during the orbital evolution, which facilitates the formation of more stable configurations and increases the probability that both planets will survive. However, the value of $e_B$ remains a determining factor for overall stability, limiting the emergence of stable systems even under more favourable migration conditions.
Low $K$ values (weak circularization with respect to migration) lead to larger eccentricities and promote instability via resonance overlap (see Sec.~\ref{subsec:Explaining_dynamics_resonant_instabilities}), while higher $K$ values enhance stability by maintaining low eccentricities.

Although the MMRs between the inner planet and the binary are not explicitly shown in Fig.~\ref{fig:impact_timescales}, we observe a clear correlation between the binary eccentricity and the degree of the resonance: systems with low $e_{B}$ are typically associated with low-degree resonances (3:1, 4:1), while higher eccentricities favour higher-degree commensurabilities (e.g. 6:1, 7:1), as discussed in \cite{Nelson_2003}. This behaviour is consistent with the results presented in Sec.~\ref{subsec:MMR_observed_BP_PP}, where $e_B$ is shown to have a stronger impact on the dynamical evolution than the binary mass ratio.

Finally, the emergence of MMRs between the planets—highlighted in the shaded areas—depends largely on the migration parameters. Since the systems are initialized near the 2:1 commensurability, capture into this resonance occurs naturally during the early stages of migration. However, whether this capture is maintained or followed by a transition to other resonances depends on the efficiency of eccentricity damping.

For efficient $e$-damping (high $K$), migration proceeds in the adiabatic regime: eccentricities remain low allowing the system to adjust its orbital configuration smoothly. In this case, although planets are initially trapped in the 2:1 MMR, continued migration can drive the system away from exact commensurability, allowing escape from resonance on a timescale $\sim \tau_e$ and subsequent capture into nearby resonances, most commonly the 3:2 MMR \citep[e.g.,][]{Goldreich_Schlichting_2014,2022MNRAS.514.3844C}, as observed in our simulations. This behaviour reflects the fact that resonance locking depends on the orbital configuration at the time of encounter rather than on the initial conditions \citep{1993Icar..103..301B}. 

In contrast, for weak eccentricity damping (low $K$), eccentricities grow significantly during migration. In this regime, although temporary capture at the 2:1 resonance still occurs, the strong eccentricity excitation typically prevents further resonant transitions. As a result, systems either become dynamically unstable or, in the stable cases, remain locked in the initial 2:1 configuration. This reflects that weak damping prevents controlled, adiabatic evolution across resonances.

Overall, this analysis shows that, while the stability of the system is largely controlled by the binary's parameters, the timescales of migration play a key role in the dynamical evolution of the planets. In particular, variations in migration, eccentricity damping, and subsequent resonance crossing determine whether the planets manage to remain in stable configurations or are driven into dynamically unstable regions, which can lead to the ejection of one or both planets and, consequently, to a reduction in the number of surviving planets. In this context, our results indicate that modifying $\tau_{a,i}$, and $K$ does not fundamentally alter the general stability trends governed by the orbital parameters of the binary, but it does affect the evolutionary trajectories followed by the CBPs as well as the range of conditions under which stable configurations can be achieved.

\section{Discussion}
\label{sec:Discussion}

\subsection{Explaining the dynamics of circumbinary planets through resonant instabilities}
\label{subsec:Explaining_dynamics_resonant_instabilities}

\begin{figure}
\centering
    \begin{tabular}{c}
        \begin{overpic}[width=0.45\textwidth]{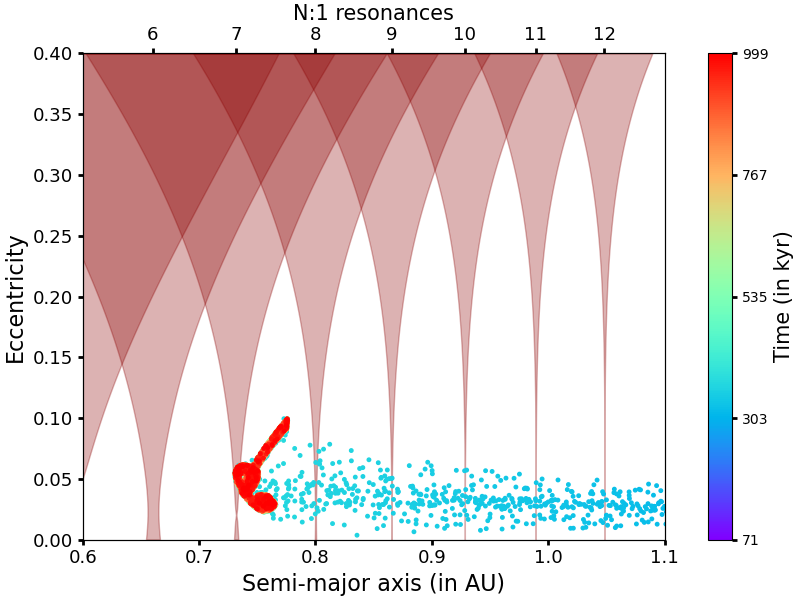}
        \put(69,35){\bfseries (a)}
        \end{overpic} \\
        \begin{overpic}[width=0.45\textwidth]{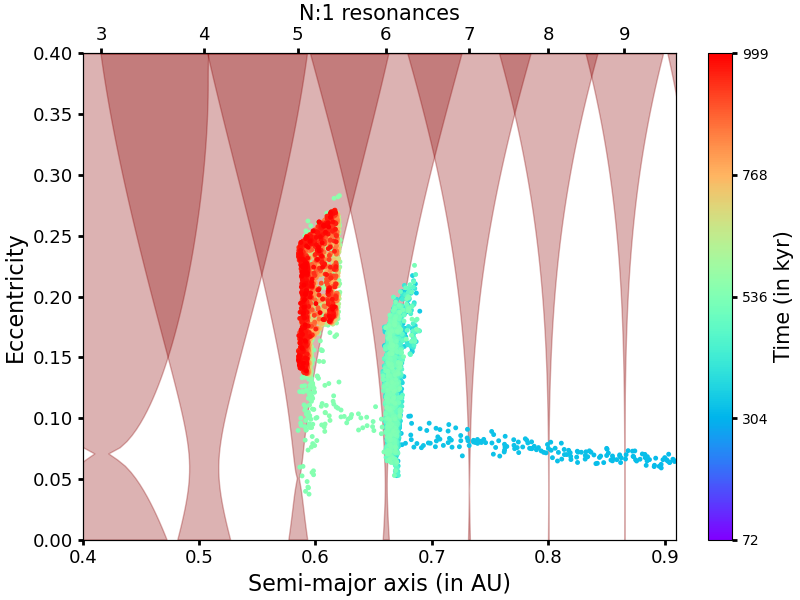}
        \put(71,35){\bfseries (b)} 
        \end{overpic} \\
        \begin{overpic}[width=0.45\textwidth]{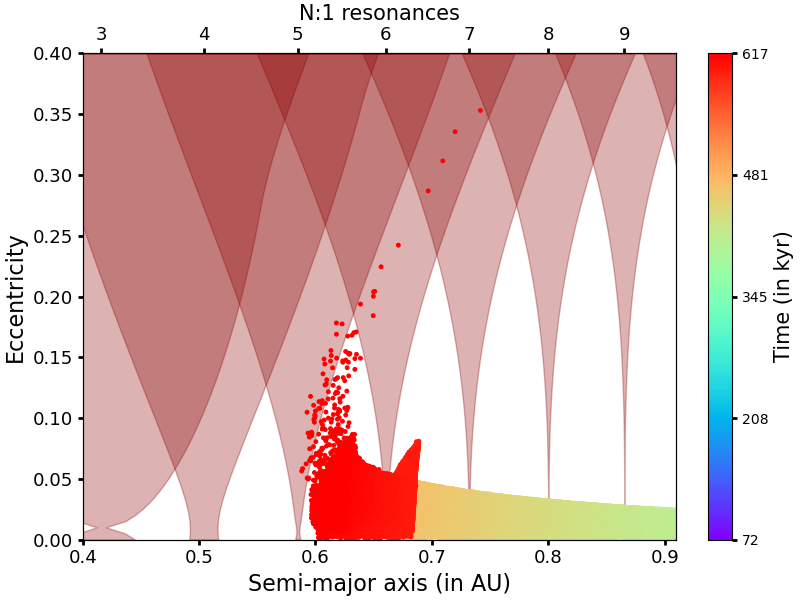}
        \put(71,35){\bfseries (c)}
        \end{overpic}
    \end{tabular}
    \caption{
        Semi-major axis versus eccentricity maps showing the evolution of one planet in three different simulations. Coloured points indicate the temporal evolution of Planet~1. Areas filled in dark red correspond to the widths of the N:1 resonances, computed following \cite{Gallardo_2021}. In all maps, Planet~1 initially remains stable at a N:1 resonance until the arrival of Planet~2, but the subsequent evolution differs: Panel~(a): Planet~1 remains stable at a 7:1 resonance; Panel~(b): Planet~1 crosses the 6:1 resonance after interacting with Planet~2 and migrates until it enters a stable 5:1 resonance; Panel~(c): Planet~1 crosses the 6:1 resonance after interacting with Planet~2 and migrates until it arrives at a 5:1 commensurability leading to its ejection.
    }
    \label{fig:a_vs_e_map}
    
\end{figure}

Our results indicate that planetary ejections in migrating circumbinary systems are primarily driven by the instability associated with N:1 resonances with the binary. Planet–planet resonances influence the dynamical evolution, but they mainly act as facilitators that enable a planet to enter resonance-overlap regions. The decisive mechanism leading to chaotic evolution and ejection is the eccentricity growth induced by N:1 resonances, which drives the planet into regions of resonance overlap.

In the $(a,e)$ parameter space, successive N:1 resonances with the binary can overlap, creating chaotic regions. If a planet enters such an overlap region, its orbit may undergo strong chaotic diffusion, potentially leading to orbit crossing, close encounters, and ejection. This phenomenon has been highlighted by \cite{Sutherland_Kratter_2019} in the case of resonant migration in circumbinary systems. However, resonance overlap typically occurs at moderate to high eccentricities. A planet with initially low eccentricity must therefore experience eccentricity growth in order to reach these unstable regions. According to \cite{Sutherland_Kratter_2019}, three mechanisms can produce this excitation: direct interaction with an N:1 resonance with the binary, planet–planet resonances, and secular forcing from the binary during migration. 

In some simulations, the planets enter a secular resonance with the binary. This is evidenced by the libration of the angles $\varpi_P - \varpi_B$, observed in several cases, which indicates that the planets are trapped in a secular resonance with the binary. Such a configuration can naturally account for the gradual growth in eccentricity observed prior to the onset of instability. Figure~\ref{app:angles_combined} illustrates an example of the evolution of the angles $\varpi_P - \varpi_B$ for both planets in the simulation presented in Figure~\ref{fig:Sim_qB0.6_eB0.5_simult}. However, secular resonances are not present in all simulations: some show no evidence of resonance, while others exhibit resonance involving either both planets or only one of them. A detailed quantitative comparison with existing analytical prescriptions \citep[e.g.,][]{Zoppetti_2019,Sutherland_Kratter_2019,Gianuzzi+2023} is not straightforward, as these works rely on different assumptions and parameter regimes. Such a comparison is beyond the scope of the present study. 

Importantly, while secular forcing contributes to the progressive increase in eccentricity, the onset of instability ultimately coincides with the system entering the N:1 resonance-overlap region, which appears to be the immediate trigger of ejection. To test whether resonance overlap is indeed responsible for planetary ejections, we analysed the evolution of the innermost planet in the $(a,e)$ plane and compared its trajectory with the location of the N:1 resonances and their associated overlap regions (see Fig.~\ref{fig:a_vs_e_map}). We find that all configurations leading to ejection show the planet entering a resonance-overlap region prior to its removal from the system. This systematic behaviour strongly indicates that chaotic diffusion associated with N:1 resonance overlap is the direct dynamical origin of the instability.
Figure~\ref{fig:a_vs_e_map} illustrates different evolutionary pathways identified in our simulations for both migration scenarios:
\begin{enumerate}
    \item In most simulations for the sequential scenario, the innermost planet is first trapped in a resonance with the binary. However, a subsequent planet–planet resonance breaks this configuration. This interaction can lead to the quasi-immediate ejection of the planet, but it can also cause further inward migration until the planet approaches the next N:1 commensurability, where eccentricity growth drives it into resonance overlap. This case is shown in Panel~(c).
    \item By contrast, in both scenarios, stable configurations can form and correspond to trajectories that remain outside the overlap regions throughout the migration (see Panel~(a)). 
    \item This happens even when the eccentricity reaches relatively large values. Panel~(b) shows a simulation similar to the one in Panel~(c), but here the innermost planet remains stable after migrating again and parking at a 5:1 commensurability. The planet is very close, but out of the overlap regions, and can remain stable until the end of the simulation.
\end{enumerate}
We also identify cases in which the planet is not permanently captured into an N:1 resonance but nevertheless undergoes strong eccentricity excitation while approaching it. This excitation alone is sufficient to drive the planet into the overlap region, ultimately leading to ejection.

Taken together, our results indicate that while planet–planet resonances shape the detailed migration history and can trigger resonance breaking, they are not the primary driver of planetary instabilities. Rather, they regulate the conditions under which a planet enters the chaotic regions generated by the overlap of N:1 resonances. Overall, the simulations that result in planetary ejections exhibit an evolutionary pathway similar to that shown in Panel~(c), providing strong evidence that resonance overlap associated with N:1 commensurabilities with the binary is the dominant mechanism controlling the stability of migrating circumbinary planets. This is consistent with the scenario proposed by \cite{Sutherland_Kratter_2019}.

It is worth noting that our simulations do not include the presence of an inner disk edge. If such an inner edge were located farther from the main binary-planet resonances, it could reduce the number of planetary ejections. In that case, a planet residing at the inner edge would avoid the dominant binary-planet resonances and their overlap, thereby preventing the large eccentricity growth that drives ejections. In this configuration, the only remaining mechanism capable of triggering close encounters would be planet-planet resonances. However, as discussed previously, these resonances are not the main cause of planetary ejections in our simulations; rather, they primarily act as a catalyst that drives the system toward the resonance-overlap regime. Therefore, we expect the planetary ejection rate to be significantly lower if migration stops at a more distant inner edge. Such a configuration could also substantially reduce, or even eliminate, the differences observed, and discussed in Section~\ref{subsec:Differences_migration_scenarios}, between the two migration scenarios.

\subsection{Differences between sequential and simultaneous migration scenarios}
\label{subsec:Differences_migration_scenarios}

In this section, we investigate the differences in stability between the two migration scenarios. Although the simultaneous scenario yields a larger number of stable two-planet systems overall (174 versus 112), our analysis does not point to a single dynamical mechanism that universally accounts for this difference. Rather, the variations in stability appear to be primarily driven by the local dynamical structure associated with the binary parameters. Nevertheless, we identify three general mechanisms that, while not universally applicable, tend to favour the formation of stable multi-planet systems in the simultaneous migration scenario.

Firstly, the difference can plausibly be attributed to the timing of eccentricity excitation associated with the 2:1 planet–planet resonance. In the simultaneous scenario, we observe the eccentricity growth occurring during resonant capture is progressively damped during the migration. In contrast, in the sequential scenario, the late arrival of Planet~2 induces additional eccentricity excitation after migration has ceased, preventing further damping. As previously said, the main stability depends on the resonance overlaps related to the binary, and even a modest increase in eccentricity may be sufficient to push the system toward instability. In other words, the simultaneous migration scenario itself acts as a mechanism that can prevent dramatic eccentricity growth for the inner planet. In contrast, in the sequential scenario, the inner planet experiences the cumulative effect of eccentricity excitations associated with resonances with both the binary and the outer planet.

Secondly, the specific choice of migration timescales may be another relevant factor. In our simulations, for both scenarios, we adopt a timescale difference of $\Delta \tau_a \approx 10^6$ yr and a ratio $\Delta \tau_a/\tau_e \approx$ 1000. As discussed in Section~\ref{sec:Methodology}, these values satisfy the criteria required to establish long-term stable MMRs between the planets. However, these conditions are only met in the simultaneous scenario. In the sequential scenario, the inner planet reaches a commensurability with the binary before entering into MMR with the outer planet. As a result, its migration stalls, and the effective timescale difference between the two planets is then set by the migration timescale of the outer planet alone. This leads to significantly smaller effective values, $\Delta \tau_a \approx 1.8 \times 10^5$ yrs and $\Delta\tau_a/\tau_e \approx$ 180 (noting that these values are in fact even smaller due to the coupling between eccentricity and semi-major axis evolution). According to the criteria of \cite{Lin+2025}, these conditions still allow for capture into MMR, but the resonance is expected to be short-lived. Interestingly, despite this expectation, we observe several simulations in the sequential scenario in which long-term MMR is maintained. Nevertheless, this mechanism likely reduces the overall probability of forming stable multi-planet systems in the sequential scenario.

Lastly, for the cases with $e_B = \{ 0.4-0.5\}$, a key difference between the two migration scenarios may be related to the role of secular resonances. Across a wide range of binary parameters, we find that planets captured in MMR with the binary exhibit significantly lower eccentricities when they remain in secular resonance after capture, compared to planets that are not in secular resonance after capture. This phenomenon mainly occurs when the longitudes of pericentre of the planet and the binary are quasi-aligned (in other words, the angle $\varpi_P - \varpi_B$ librates around 0°). Secular resonances are not needed in all binaries to form multi-planet systems: several cases with low binary eccentricity produce stable multi-planet systems with no planet in secular resonance. However, in all simulations with $e_B \ge 0.4$ and leading to stable multi-planet systems, the two planets remain in secular resonance after the resonant captures. Therefore, the fact that both planets remain in secular resonance with the binary may constitute a favourable condition for the survival of multi-planet systems around highly eccentric binaries. Indeed, in the absence of secular resonances, the eccentricity excitation may become large enough for the system to enter resonance overlap regions, leading to the ejection of one planet. Even though the sequential scenario can generate planets in secular resonance, we observe only 4 simulations in which both planets remain stable in secular resonance, contrary to 35 in the simultaneous scenario.

 \begin{figure}
  \centering
  \begin{minipage}{0.45\textwidth}
    \includegraphics[width=1\textwidth]{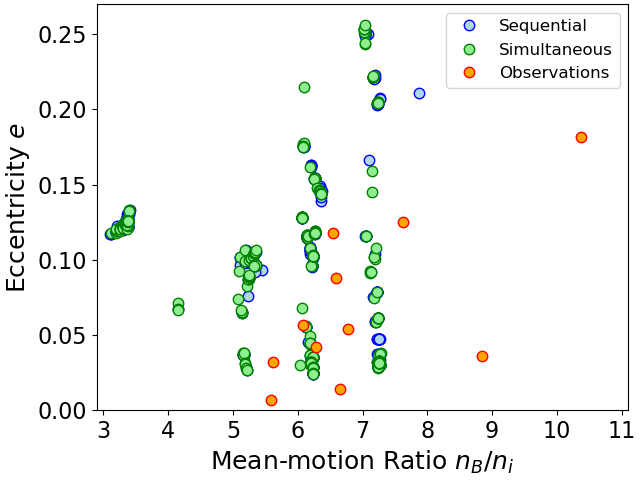}
  \end{minipage}
  \caption{
    Distribution of the eccentricity $e$ as a function of the mean-motion ratio $n_B/n_i$ for all circumbinary planets that are the sole survivors in our simulations (i.e., in initially two-planet systems, we only consider cases where a single planet remains at the end of the integration). Blue points correspond to planets obtained in the sequential migration scenario, while green points represent planets produced in the simultaneous migration scenario. Orange points indicate the observed circumbinary planets that are single in their systems, orbit main-sequence binaries, and have mean-motion ratios $n_B/n_i$ between 3 and 11.
  }
  \label{fig:e_vs_P_plot}
\end{figure}

\subsection{Impact of migration timescales, planetary masses, and limitations}
\label{subsec:impact_migration_timescales_masses}

In our simulations, we adopt migration rates consistent with those expected for 10‑Earth-mass planets in the inner regions of a disc with specific surface density and viscosity, following \cite{Cresswell_Nelson2008}. However, our results contrast with those reported in several studies of N:1 resonances in circumbinary systems, such as \cite{Martin_Fitzmaurice2022} and \cite{Gianuzzi+2023}. While, in our parameter space, planets captured into MMR with the binary typically remain stable, these studies report the opposite behaviour, with the vast majority of planets being ejected shortly after entering resonance.
Although the migration models adopted in these works differ from ours, we argue that the primary source of discrepancy lies in the value of the ratio $K = \tau_a/\tau_e$. In both \cite{Martin_Fitzmaurice2022} and \cite{Gianuzzi+2023}, a typical value of $K=10$ is used, following earlier studies \citep{Lee_Peale_2002, Kley_Peitz_Bryden_2004}. This value is significantly lower than those explored in our suite of 1200 simulations. As shown in Figure~\ref{fig:impact_timescales}, increasing $K$ generally enhances the stability of planetary systems. To test this hypothesis, we reproduced the simulations of \cite{Martin_Fitzmaurice2022} using their migration model, and then repeated them with an increased value of $K$ at 100. We find that, in this regime, circumbinary planets can be captured into MMR with the binary and remain stable over much longer timescales compared to the $K=10$ case.

Several Type I planetary migration models \cite[e.g.][]{Cresswell_Nelson2008, Ida+2020} provide prescriptions for $\tau_a$ and $\tau_e$ that indicate the ratio $K$ is not fixed at a value of 10 (instead, it strongly depends on parameters like the disc aspect ratio $h/r$). While disc conditions leading to $K \approx$ 10 cannot be excluded, typical disc parameters are expected to yield significantly larger values, which can substantially affect planetary stability. In light of these considerations, although our approach does not represent the most realistic migration framework, our results suggest that the commonly held view that MMRs with the binary act primarily as a barrier to the stability of circumbinary planets should be treated with caution. Instead, stable capture into such resonances appears to be entirely feasible, with the long-term stability of these configurations depending sensitively on the properties of the protoplanetary disc through which the planets migrate.

The values of $\tau_a$ and $\tau_e$ also depend on both the stellar and planetary masses. According to \cite{Cresswell_Nelson2008}, for a fixed binary mass, decreasing the planetary mass leads to an increase in $\tau_a$. Based on the results shown in Figure~\ref{fig:impact_timescales}, increasing $\tau_a$ appears to favour the formation of stable multi-planet systems. This suggests that lower-mass planets (a few $M_{\oplus}$) may be more prone to forming stable multi-planet configurations.
However, only two circumbinary planets with masses below 10 $M_{\oplus}$ have been observed so far: Kepler-47b and Kepler-47c \citep[see][]{Orosz_2019}. Another factor that can influence stability is the mass ratio between the two planets. In our study, we consider the simplified case where both planets have equal masses, fixed at 10 $M_{\oplus}$. Yet, as shown by \cite{Lin+2025}, the conditions required to establish a stable resonance between the planets depend strongly on this mass ratio. When the inner planet is significantly more massive than the outer one, and the planets converge slowly, stable MMR capture can occur relatively easily, even for small values of $\Delta\tau_a/\tau_e$. In contrast, when the outer planet is more massive than the inner one, the conditions required to form a stable MMR become more restrictive than in the equal-mass case. One would therefore expect planetary architectures in which the more massive planets are located closer to the binary. However, the two known circumbinary multi-planet systems do not appear to support this trend.

Several factors may explain these discrepancies. Our study considered only a simplified case of two 10‑Earth-mass planets, whereas the observed systems differ in planet number and masses. The chaotic nature of the dynamics implies that such variations can strongly influence outcomes. Furthermore, our migration model is highly simplified, as it assumes a constant migration rate independent of the planet's position. However, we argued that the most critical factor is the value of the migration timescales at the locations where resonances are expected to occur.

To assess the robustness of our results, we performed an additional set of simulations in the sequential scenario ($a_{P_1} = 6$ ~AU; $a_{P_2} = 30$~AU) using the migration prescription of \cite{Cresswell_Nelson2008} and the disk parameters described in Section~\ref{subsec:Migration model}. We considered nine combinations of binary parameters, with $q_B \in \{0.1, 0.5, 1.0\}$ and $e_B \in \{0.0, 0.2, 0.4\}$. Since the migration rate depends on the planet's position, migration proceeds significantly more slowly than in our previous simulations. Consequently, the integration time was extended to 8 Myr. The results, shown in Figure~\ref{fig:mosaic_migration_tests}, indicate that the resonant configurations and parking locations identified in our nominal simulations are generally recovered when a more realistic migration prescription is adopted. However, we note differences in the survival rates, which are generally lower than in our nominal case. This leads to more pessimistic estimates for circumbinary planet stability. For example, in the case $q_B = 0.1$ and $e_B = 0.4$, none of the simulations retained a stable planet, whereas all ten simulations using a constant migration timescale resulted in the survival of one planet. One possible explanation for this discrepancy is that planets located near resonance do not share identical migration timescales. Indeed, the timescales computed in Section 2.2 assume circular planetary orbits. In practice, planetary eccentricities typically reach values between 0.04 and 0.06. This naturally leads to different migration rates as the prescription of \cite{Cresswell_Nelson2008} is highly sensitive to eccentricity.

Another factor that may contribute to the observed differences is the possibility of migration reversal, where inward migration transitions into outward migration. In simulations with $e_B = 0.0$ and $q_B \in \{0.5, 1.0\}$, we observed that the inner planet halted its migration near a period ratio of $4$ without exhibiting libration of the resonant angles associated with the $4:1$ mean-motion resonance. Moreover, this stopping process was considerably more gradual than the abrupt halting typically associated with resonance capture. A plausible explanation is that the migration torques acting on the planet effectively cancelled each other, preventing further inward migration before the planet could reach the expected $3:1$ resonance. Overall, these additional simulations do not challenge our main conclusions: planets can be trapped in resonances, their survival strongly depends on the binary parameters, and migration timescales play a crucial role in determining the final system architecture. Nevertheless, contrary to the trend identified in Section 3.4, we find that larger values of $\tau_a$ and $\tau_a/\tau_e$ do not necessarily guarantee stability. This suggests that the relationship between migration efficiency and planetary survival is more complex than the one predicted by our simplified model.

Additional assumptions were made in the disk model including the absence of disc structure, thermodynamic effects or planetary accretion. It is important to emphasize that we assume a circular protoplanetary disk, even for highly eccentric binaries. This approximation neglects the fact that the binary can affect the inner regions of the disk, modify its density profile, and consequently affect planetary migration. Indeed, several studies \citep[e.g.,][]{Hirsh_2020,Ragusa_2020} have shown that the inner disk region can become eccentric, even for circular binaries, and that the eccentricity of the gap increases with the binary eccentricity. In some cases, such as the planet migration simulations of the Kepler-34 system presented by \cite{Thun_Kley2018}, the disk inner edge is located well inside the resonances identified in our study. Despite the presence of an eccentric disk, planets are therefore able to migrate inward and reach the disk inner edge. Interestingly, \cite{Nelson_2003} identified a particular evolutionary mode for binaries with ($e_{\rm bin} \geq 0.2$). The interaction between an eccentric disk and the planet can, in some cases, reduce the net torque driving migration, or even reverse its sign. This mechanism can therefore halt planetary migration before the planet enters the inner disk region. These studies show that eccentric discs do not prevent the survival of circumbinary planets, but they can lead to planetary locations that differ from those found in this work.

Type-I migration favours radial drift, making convergent migration difficult for equal-mass planets; sustained convergence requires a much less massive inner planet, which is itself challenging to achieve \citep{Fitzmaurice+2022}. Also, the values of $\tau_a$ and $\tau_e$ have been chosen independently of the planetary masses. However, the masses of planets formed in the disc remain linked to the initial disc conditions, which themselves determine the migration timescales for each planet. The stability maps should therefore be interpreted with this aspect in mind. Finally, other potentially important effects not included here are an inner disc cavity, tidal interactions with the binary \citep{Zoppetti_2020}, and long-term binary evolution such as contraction and circularization. 

Lastly, our case study focused on the early dynamical history of circumbinary planets, namely migration within a protoplanetary disc. Later evolutionary processes may further affect planetary stability. Several million years after disc formation, photoevaporation can disperse the disc from the inside out, potentially perturbing compact, otherwise stable configurations \citep[e.g., the Solar case studied in][]{Liu_2022}. Even after disc dispersal, gravitational interactions between planets or with the central binary can modify the system, either on short timescales if planets are closely spaced, or over long timescales through secular perturbations. Additionally, the binary itself may evolve, through tidal shrinkage or circularization, which could further reshape the planetary dynamics (Sucerquia, Cuello \& Duchêne, in prep.). Therefore, our results indicate which binary configurations do not prevent the formation of circumbinary planetary systems during migration, rather than identifying which binaries are more likely to host these systems in general. In Figure~\ref{fig:e_vs_P_plot}, we show all single surviving planets at the end of our simulations in the eccentricity–orbital period plane, and compare them to observed single-planet systems. Several observed planets lie within the region of MMRs with the binary. However, most of them have been detected either between the predicted regions or beyond these. This likely provides a strong indication of the importance of processes occurring after the planet migration phase.

\section{Conclusions}
\label{sec:Conclusions}

In this study, we performed N-body simulations of two circumbinary planets undergoing inward migration to investigate how the binary parameters shape the resulting final architectures. We considered two distinct migration scenarios: the sequential scenario, in which the planets migrate independently, and the simultaneous scenario, in which both planets migrate together while locked in a 2:1 MMR. Our main findings are the following:

   \begin{itemize}
      \item A circumbinary planet can halt its migration by being captured into a MMR with the binary and, in contrast to what has been reported in several previous studies, remain stable over long timescales. This behaviour is observed across a wide range of parameter space and provides an additional possible parking location, alongside the inner edge of the circumbinary disc.
      \item The formation of stable multi-planet systems is also possible; however, this stability strongly depends on the binary parameters $q_B$ and $e_B$. We also find a strong dependence of planetary stability on the migration timescales $\tau_a$ and $\tau_e$, which themselves are functions of both the disc properties and the planet’s characteristics.
      \item In general, the formation of multi-planet systems is more favourable when planets migrate in the simultaneous scenario than in the sequential one. This trend is particularly pronounced for highly eccentric binaries ($e_B$ = 0.4–0.5).
   \end{itemize}

Among currently known circumbinary systems, only Kepler-47 \citep{Orosz_2012} and TOI-1338 \citep{Kostov_2020,Standing+2023} host multiple planets. Using their measured binary mass ratios and eccentricities ($q_B$ = 0.357, $e_B$ = 0.029 for Kepler-47; $q_B$ = 0.281, $e_B$ = 0.155 for TOI-1338), we placed these systems on our stability maps and compared their locations with the predicted stable and unstable regions. Contrary to expectations, their positions do not systematically coincide with regions that favour fully stable two-planet configurations. This could be related to several limitations of our simulations, such as the absence of an inner disc edge, or the fact that we only model the migration phase, without accounting for subsequent evolutionary stages such as disc photoevaporation.

Despite these limitations, our work highlights the complexity of circumbinary multi-planet stability and identifies broad trends across parameter space. Future studies should refine the stability regions using more realistic migration models and including additional physical processes. Moreover, we assumed that the disk and the binary were coplanar. However, misaligned circumbinary disks are known to exist \citep[e.g.,][]{Chiang_Murray-Clay_2004,Kennedy_2019, Cuello+2025}. In such systems, additional dynamical effects, such as Kozai-Lidov effects, may play an important role and could significantly alter the process of resonant capture, either with the binary or with a companion planet. While \cite{Chen+2023} investigated the long-term stability of misaligned circumbinary planets, they did not consider the migration phase within a misaligned protoplanetary disk. Extending our analysis to a broader range of system architectures and parameters, including orbital inclinations, planet multiplicity, planetary masses, and total binary mass, would provide a more comprehensive understanding of the formation and long-term stability of circumbinary planetary systems. Ultimately, these stability maps may guide future observational campaigns, helping to identify promising binaries for low-mass multi-planet circumbinary systems. These upcoming surveys will allow us to establish the dominant mechanisms affecting long-term stability of such architectures.

\begin{acknowledgements}
    We kindly thank the anonymous referee for their constructive comments and feedback. This project has received funding from the European Research Council (ERC) under the European Union Horizon Europe programme (grant agreement No. 101042275, project Stellar-MADE). MAB acknowledges support from Agencia Nacional de Investigación y Desarrollo (ANID) through FONDECYT Regular n$^\circ$1262342. CC acknowledges support from ANID through FONDECYT Postdoctoral grant n$^\circ$3230283 and CAS-ANID (CASSACA) n$^\circ$250005. This research was also supported by the ERC/UKRI Frontier Research Guarantee programme (EP/Z000327/1/CandY). We also thank David V. Martin for his valuable suggestions, which helped us improve the methodology and interpretation of our results.
    
\end{acknowledgements}

\bibliographystyle{aa} 
\bibliography{biblio} 

@ARTICLE{Pierens+2020,
       author = {{Pierens}, Arnaud and {McNally}, Colin P. and {Nelson}, Richard P.},
        title = "{Hydrodynamical turbulence in eccentric circumbinary discs and its impact on the in situ formation of circumbinary planets}",
      journal = {\mnras},
         year = 2020,
        month = aug,
       volume = {496},
       number = {3},
        pages = {2849-2867},
          doi = {10.1093/mnras/staa1550},
archivePrefix = {arXiv},
       eprint = {2005.14693},
 primaryClass = {astro-ph.EP},
       adsurl = {https://ui.adsabs.harvard.edu/abs/2020MNRAS.496.2849P}
}

@ARTICLE{Cuello+2025,
       author = {{Cuello}, Nicol{\'a}s and {Alaguero}, Antoine and {Poblete}, Pedro P.},
        title = "{Circumstellar and Circumbinary Discs in Multiple Stellar Systems}",
      journal = {Symmetry},
         year = 2025,
        month = feb,
       volume = {17},
       number = {3},
          eid = {344},
        pages = {344},
          doi = {10.3390/sym17030344},
archivePrefix = {arXiv},
       eprint = {2501.19249},
 primaryClass = {astro-ph.EP},
       adsurl = {https://ui.adsabs.harvard.edu/abs/2025Symm...17..344C}
}

@ARTICLE{Doyle+2011,
       author = {{Doyle}, Laurance R. and {Carter}, Joshua A. and {Fabrycky}, Daniel C. and {Slawson}, Robert W. and {Howell}, Steve B. and {Winn}, Joshua N. and {Orosz}, Jerome A. and {P{\v{r}}sa}, Andrej and {Welsh}, William F. and {Quinn}, Samuel N. and {Latham}, David and {Torres}, Guillermo and {Buchhave}, Lars A. and {Marcy}, Geoffrey W. and {Fortney}, Jonathan J. and {Shporer}, Avi and {Ford}, Eric B. and {Lissauer}, Jack J. and {Ragozzine}, Darin and {Rucker}, Michael and {Batalha}, Natalie and {Jenkins}, Jon M. and {Borucki}, William J. and {Koch}, David and {Middour}, Christopher K. and {Hall}, Jennifer R. and {McCauliff}, Sean and {Fanelli}, Michael N. and {Quintana}, Elisa V. and {Holman}, Matthew J. and {Caldwell}, Douglas A. and {Still}, Martin and {Stefanik}, Robert P. and {Brown}, Warren R. and {Esquerdo}, Gilbert A. and {Tang}, Sumin and {Furesz}, Gabor and {Geary}, John C. and {Berlind}, Perry and {Calkins}, Michael L. and {Short}, Donald R. and {Steffen}, Jason H. and {Sasselov}, Dimitar and {Dunham}, Edward W. and {Cochran}, William D. and {Boss}, Alan and {Haas}, Michael R. and {Buzasi}, Derek and {Fischer}, Debra},
        title = "{Kepler-16: A Transiting Circumbinary Planet}",
      journal = {Science},
         year = 2011,
        month = sep,
       volume = {333},
       number = {6049},
        pages = {1602},
          doi = {10.1126/science.1210923},
archivePrefix = {arXiv},
       eprint = {1109.3432},
 primaryClass = {astro-ph.EP},
       adsurl = {https://ui.adsabs.harvard.edu/abs/2011Sci...333.1602D}
}

@ARTICLE{Standing+2023,
       author = {{Standing}, Matthew R. and {Sairam}, Lalitha and {Martin}, David V. and {Triaud}, Amaury H.~M.~J. and {Correia}, Alexandre C.~M. and {Coleman}, Gavin A.~L. and {Baycroft}, Thomas A. and {Kunovac}, Vedad and {Boisse}, Isabelle and {Cameron}, Andrew Collier and {Dransfield}, Georgina and {Faria}, Jo{\~a}o P. and {Gillon}, Micha{\"e}l and {Hara}, Nathan C. and {Hellier}, Coel and {Howard}, Jonathan and {Lane}, Ellie and {Mardling}, Rosemary and {Maxted}, Pierre F.~L. and {Miller}, Nicola J. and {Nelson}, Richard P. and {Orosz}, Jerome A. and {Pepe}, Franscesco and {Santerne}, Alexandre and {Sebastian}, Daniel and {Udry}, St{\'e}phane and {Welsh}, William F.},
        title = "{Radial-velocity discovery of a second planet in the TOI-1338/BEBOP-1 circumbinary system}",
      journal = {Nature Astronomy},
         year = 2023,
        month = jun,
       volume = {7},
        pages = {702-714},
          doi = {10.1038/s41550-023-01948-4},
archivePrefix = {arXiv},
       eprint = {2301.10794},
 primaryClass = {astro-ph.EP},
       adsurl = {https://ui.adsabs.harvard.edu/abs/2023NatAs...7..702S}
}

@ARTICLE{Baycroft+2025,
       author = {{Baycroft}, Thomas A. and {Santerne}, Alexandre and {Triaud}, Amaury H.~M.~J. and {Heidari}, Neda and {Sebastian}, Daniel and {Davis}, Yasmin T. and {Correia}, Alexandre C.~M. and {Sairam}, Lalitha and {Freckelton}, Alix V. and {Adamson}, Aleyna and {Boisse}, Isabelle and {Coleman}, Gavin A.~L. and {Dransfield}, Georgina and {Faria}, Jo{\~a}o and {Grouffal}, Salom{\'e} and {Hara}, Nathan and {H{\'e}brard}, Guillaume and {Kunovac}, Vedad and {Martin}, David V. and {Maxted}, Pierre F.~L. and {Nelson}, Richard P. and {Scott}, Madison G. and {Scutt}, Owen J. and {Standing}, Matthew R.},
        title = "{BEBOP VII. SOPHIE discovery of BEBOP-3b, a circumbinary giant planet on an eccentric orbit}",
      journal = {\mnras},
         year = 2025,
        month = aug,
       volume = {541},
       number = {3},
        pages = {2801-2814},
          doi = {10.1093/mnras/staf1184},
archivePrefix = {arXiv},
       eprint = {2506.14615},
 primaryClass = {astro-ph.EP},
       adsurl = {https://ui.adsabs.harvard.edu/abs/2025MNRAS.541.2801B}
}

@ARTICLE{Goldberg+2023,
       author = {{Goldberg}, Max and {Fabrycky}, Daniel and {Martin}, David V. and {Albrecht}, Simon and {Deeg}, Hans J. and {Nowak}, Grzegorz},
        title = "{A 5M$_{Jup}$ non-transiting coplanar circumbinary planet around Kepler-1660AB}",
      journal = {\mnras},
         year = 2023,
        month = nov,
       volume = {525},
       number = {3},
        pages = {4628-4641},
          doi = {10.1093/mnras/stad2568},
archivePrefix = {arXiv},
       eprint = {2308.09255},
 primaryClass = {astro-ph.EP},
       adsurl = {https://ui.adsabs.harvard.edu/abs/2023MNRAS.525.4628G}
}

@ARTICLE{Holman_Wiegert1999,
       author = {{Holman}, Matthew J. and {Wiegert}, Paul A.},
        title = "{Long-Term Stability of Planets in Binary Systems}",
      journal = {\aj},
         year = 1999,
        month = jan,
       volume = {117},
       number = {1},
        pages = {621-628},
          doi = {10.1086/300695},
archivePrefix = {arXiv},
       eprint = {astro-ph/9809315},
 primaryClass = {astro-ph},
       adsurl = {https://ui.adsabs.harvard.edu/abs/1999AJ....117..621H}
}

@ARTICLE{Meschiari2012,
       author = {{Meschiari}, Stefano},
        title = "{Planet Formation in Circumbinary Configurations: Turbulence Inhibits Planetesimal Accretion}",
      journal = {\apjl},
         year = 2012,
        month = dec,
       volume = {761},
       number = {1},
          eid = {L7},
        pages = {L7},
          doi = {10.1088/2041-8205/761/1/L7},
archivePrefix = {arXiv},
       eprint = {1210.7757},
 primaryClass = {astro-ph.EP},
       adsurl = {https://ui.adsabs.harvard.edu/abs/2012ApJ...761L...7M}
}

@ARTICLE{Paardekooper+2012,
       author = {{Paardekooper}, Sijme-Jan and {Leinhardt}, Zo{\"e} M. and {Th{\'e}bault}, Philippe and {Baruteau}, Cl{\'e}ment},
        title = "{How Not to Build Tatooine: The Difficulty of In Situ Formation of Circumbinary Planets Kepler 16b, Kepler 34b, and Kepler 35b}",
      journal = {\apjl},
         year = 2012,
        month = jul,
       volume = {754},
       number = {1},
          eid = {L16},
        pages = {L16},
          doi = {10.1088/2041-8205/754/1/L16},
archivePrefix = {arXiv},
       eprint = {1206.3484},
 primaryClass = {astro-ph.EP},
       adsurl = {https://ui.adsabs.harvard.edu/abs/2012ApJ...754L..16P}
}

@ARTICLE{Pierens_Nelson2008a,
       author = {{Pierens}, A. and {Nelson}, R.~P.},
        title = "{On the evolution of multiple low mass planets embedded in a circumbinary disc}",
      journal = {\aap},
         year = 2008,
        month = feb,
       volume = {478},
       number = {3},
        pages = {939-949},
          doi = {10.1051/0004-6361:20078844},
archivePrefix = {arXiv},
       eprint = {0712.0961},
 primaryClass = {astro-ph},
       adsurl = {https://ui.adsabs.harvard.edu/abs/2008A&A...478..939P}
}

@ARTICLE{Pierens_Nelson2008b,
       author = {{Pierens}, A. and {Nelson}, R.~P.},
        title = "{On the formation and migration of giant planets in circumbinary discs}",
      journal = {\aap},
         year = 2008,
        month = may,
       volume = {483},
       number = {2},
        pages = {633-642},
          doi = {10.1051/0004-6361:200809453},
archivePrefix = {arXiv},
       eprint = {0803.2000},
 primaryClass = {astro-ph},
       adsurl = {https://ui.adsabs.harvard.edu/abs/2008A&A...483..633P}
}

@ARTICLE{Kley_Haghighipour2014,
       author = {{Kley}, Wilhelm and {Haghighipour}, Nader},
        title = "{Modeling circumbinary planets: The case of Kepler-38}",
      journal = {\aap},
         year = 2014,
        month = apr,
       volume = {564},
          eid = {A72},
        pages = {A72},
          doi = {10.1051/0004-6361/201323235},
archivePrefix = {arXiv},
       eprint = {1401.7648},
 primaryClass = {astro-ph.EP},
       adsurl = {https://ui.adsabs.harvard.edu/abs/2014A&A...564A..72K}
}

@ARTICLE{Kley+2019,
       author = {{Kley}, Wilhelm and {Thun}, Daniel and {Penzlin}, Anna B.~T.},
        title = "{Circumbinary discs with radiative cooling and embedded planets}",
      journal = {\aap},
         year = 2019,
        month = jul,
       volume = {627},
          eid = {A91},
        pages = {A91},
          doi = {10.1051/0004-6361/201935503},
archivePrefix = {arXiv},
       eprint = {1905.08631},
 primaryClass = {astro-ph.EP},
       adsurl = {https://ui.adsabs.harvard.edu/abs/2019A&A...627A..91K}
}

@ARTICLE{Penzlin+2021,
       author = {{Penzlin}, Anna B.~T. and {Kley}, Wilhelm and {Nelson}, Richard P.},
        title = "{Parking planets in circumbinary discs}",
      journal = {\aap},
         year = 2021,
        month = jan,
       volume = {645},
          eid = {A68},
        pages = {A68},
          doi = {10.1051/0004-6361/202039319},
archivePrefix = {arXiv},
       eprint = {2012.03651},
 primaryClass = {astro-ph.EP},
       adsurl = {https://ui.adsabs.harvard.edu/abs/2021A&A...645A..68P}
}

@ARTICLE{Rein_Liu2012,
       author = {{Rein}, H. and {Liu}, S.-F.},
        title = "{REBOUND: an open-source multi-purpose N-body code for collisional dynamics}",
      journal = {\aap},
         year = 2012,
        month = jan,
       volume = {537},
          eid = {A128},
        pages = {A128},
          doi = {10.1051/0004-6361/201118085},
archivePrefix = {arXiv},
       eprint = {1110.4876},
 primaryClass = {astro-ph.EP},
       adsurl = {https://ui.adsabs.harvard.edu/abs/2012A&A...537A.128R}
}

@ARTICLE{Rein_Spiegel2015,
       author = {{Rein}, Hanno and {Spiegel}, David S.},
        title = "{IAS15: a fast, adaptive, high-order integrator for gravitational dynamics, accurate to machine precision over a billion orbits}",
      journal = {\mnras},
         year = 2015,
        month = jan,
       volume = {446},
       number = {2},
        pages = {1424-1437},
          doi = {10.1093/mnras/stu2164},
archivePrefix = {arXiv},
       eprint = {1409.4779},
 primaryClass = {astro-ph.EP},
       adsurl = {https://ui.adsabs.harvard.edu/abs/2015MNRAS.446.1424R}
}

@ARTICLE{Cresswell_Nelson2008,
       author = {{Cresswell}, P. and {Nelson}, R.~P.},
        title = "{Three-dimensional simulations of multiple protoplanets embedded in a protostellar disc}",
      journal = {\aap},
         year = 2008,
        month = may,
       volume = {482},
       number = {2},
        pages = {677-690},
          doi = {10.1051/0004-6361:20079178},
archivePrefix = {arXiv},
       eprint = {0811.4322},
 primaryClass = {astro-ph},
       adsurl = {https://ui.adsabs.harvard.edu/abs/2008A&A...482..677C}
}

@ARTICLE{Chen+2023,
       author = {{Chen}, Cheng and {Lubow}, Stephen H. and {Martin}, Rebecca G. and {Nixon}, C.~J.},
        title = "{Orbital stability of two circumbinary planets around misaligned eccentric binaries}",
      journal = {\mnras},
         year = 2023,
        month = jun,
       volume = {521},
       number = {4},
        pages = {5033-5045},
          doi = {10.1093/mnras/stad739},
archivePrefix = {arXiv},
       eprint = {2303.05379},
 primaryClass = {astro-ph.EP},
       adsurl = {https://ui.adsabs.harvard.edu/abs/2023MNRAS.521.5033C}
}

@ARTICLE{Gianuzzi+2023,
       author = {{Gianuzzi}, Emmanuel and {Giuppone}, Cristian and {Cuello}, Nicol{\'a}s},
        title = "{Circumbinary planets: migration, trapping in mean-motion resonances, and ejection}",
      journal = {\aap},
         year = 2023,
        month = jan,
       volume = {669},
          eid = {A123},
        pages = {A123},
          doi = {10.1051/0004-6361/202244902},
archivePrefix = {arXiv},
       eprint = {2211.08520},
 primaryClass = {astro-ph.EP},
       adsurl = {https://ui.adsabs.harvard.edu/abs/2023A&A...669A.123G}
}

@ARTICLE{Pierens_Nelson2013,
       author = {{Pierens}, A. and {Nelson}, R.~P.},
        title = "{Migration and gas accretion scenarios for the Kepler 16, 34, and 35 circumbinary planets}",
      journal = {\aap},
         year = 2013,
        month = aug,
       volume = {556},
          eid = {A134},
        pages = {A134},
          doi = {10.1051/0004-6361/201321777},
archivePrefix = {arXiv},
       eprint = {1307.0713},
 primaryClass = {astro-ph.EP},
       adsurl = {https://ui.adsabs.harvard.edu/abs/2013A&A...556A.134P}
}

@ARTICLE{Mutter+2017,
       author = {{Mutter}, Matthew M. and {Pierens}, Arnaud and {Nelson}, Richard P.},
        title = "{The role of disc self-gravity in circumbinary planet systems - II. Planet evolution}",
      journal = {\mnras},
         year = 2017,
        month = aug,
       volume = {469},
       number = {4},
        pages = {4504-4522},
          doi = {10.1093/mnras/stx1113},
archivePrefix = {arXiv},
       eprint = {1705.03035},
 primaryClass = {astro-ph.EP},
       adsurl = {https://ui.adsabs.harvard.edu/abs/2017MNRAS.469.4504M}
}

@ARTICLE{Thun_Kley2018,
       author = {{Thun}, Daniel and {Kley}, Wilhelm},
        title = "{Migration of planets in circumbinary discs}",
      journal = {\aap},
         year = 2018,
        month = aug,
       volume = {616},
          eid = {A47},
        pages = {A47},
          doi = {10.1051/0004-6361/201832804},
archivePrefix = {arXiv},
       eprint = {1806.00314},
 primaryClass = {astro-ph.EP},
       adsurl = {https://ui.adsabs.harvard.edu/abs/2018A&A...616A..47T}
}

@ARTICLE{Zoppetti+2018,
       author = {{Zoppetti}, F.~A. and {Beaug{\'e}}, C. and {Leiva}, A.~M.},
        title = "{Resonant capture and tidal evolution in circumbinary systems: testing the case of Kepler-38}",
      journal = {\mnras},
         year = 2018,
        month = jul,
       volume = {477},
       number = {4},
        pages = {5301-5311},
          doi = {10.1093/mnras/sty1002},
       adsurl = {https://ui.adsabs.harvard.edu/abs/2018MNRAS.477.5301Z}
}

@ARTICLE{Martin_Fitzmaurice2022,
       author = {{Martin}, David V. and {Fitzmaurice}, Evan},
        title = "{Running the gauntlet - survival of small circumbinary planets migrating through destabilizing resonances}",
      journal = {\mnras},
         year = 2022,
        month = may,
       volume = {512},
       number = {1},
        pages = {602-616},
          doi = {10.1093/mnras/stac090},
archivePrefix = {arXiv},
       eprint = {2112.00786},
 primaryClass = {astro-ph.EP},
       adsurl = {https://ui.adsabs.harvard.edu/abs/2022MNRAS.512..602M}
}

@ARTICLE{Fitzmaurice+2022,
       author = {{Fitzmaurice}, Evan and {Martin}, David V. and {Fabrycky}, Daniel C.},
        title = "{Sculpting the circumbinary planet size distribution through resonant interactions with companion planets}",
      journal = {\mnras},
         year = 2022,
        month = jun,
       volume = {512},
       number = {4},
        pages = {5023-5036},
          doi = {10.1093/mnras/stac741},
archivePrefix = {arXiv},
       eprint = {2202.11719},
 primaryClass = {astro-ph.EP},
       adsurl = {https://ui.adsabs.harvard.edu/abs/2022MNRAS.512.5023F}
}

@ARTICLE{Kley_Haghighipour2015,
       author = {{Kley}, Wilhelm and {Haghighipour}, Nader},
        title = "{Evolution of circumbinary planets around eccentric binaries: The case of Kepler-34}",
      journal = {\aap},
         year = 2015,
        month = sep,
       volume = {581},
          eid = {A20},
        pages = {A20},
          doi = {10.1051/0004-6361/201526648},
archivePrefix = {arXiv},
       eprint = {1506.07026},
 primaryClass = {astro-ph.EP},
       adsurl = {https://ui.adsabs.harvard.edu/abs/2015A&A...581A..20K}
}

@ARTICLE{Thebault_Bonanni2025,
       author = {{Thebault}, P. and {Bonanni}, D.},
        title = "{A complete census of planet-hosting binaries}",
      journal = {\aap},
         year = 2025,
        month = aug,
       volume = {700},
          eid = {A106},
        pages = {A106},
          doi = {10.1051/0004-6361/202555457},
archivePrefix = {arXiv},
       eprint = {2506.18759},
 primaryClass = {astro-ph.EP},
       adsurl = {https://ui.adsabs.harvard.edu/abs/2025A&A...700A.106T}
}

@ARTICLE{Martin_Triaud2014,
       author = {{Martin}, David V. and {Triaud}, Amaury H.~M.~J.},
        title = "{Planets transiting non-eclipsing binaries}",
      journal = {\aap},
         year = 2014,
        month = oct,
       volume = {570},
          eid = {A91},
        pages = {A91},
          doi = {10.1051/0004-6361/201323112},
archivePrefix = {arXiv},
       eprint = {1404.5360},
 primaryClass = {astro-ph.EP},
       adsurl = {https://ui.adsabs.harvard.edu/abs/2014A&A...570A..91M}
}

@ARTICLE{Lin+2025,
       author = {{Lin}, Linghong and {Liu}, Beibei and {Zheng}, Zekai},
        title = "{Resonance capture and stability analysis for planet pairs under Type I disk migration}",
      journal = {\aap},
         year = 2025,
        month = oct,
       volume = {702},
          eid = {A161},
        pages = {A161},
          doi = {10.1051/0004-6361/202453589},
archivePrefix = {arXiv},
       eprint = {2501.12650},
 primaryClass = {astro-ph.EP},
       adsurl = {https://ui.adsabs.harvard.edu/abs/2025A&A...702A.161L}
}

@ARTICLE{Coleman_Nelson2014,
       author = {{Coleman}, Gavin A.~L. and {Nelson}, Richard P.},
        title = "{On the formation of planetary systems via oligarchic growth in thermally evolving viscous discs}",
      journal = {\mnras},
         year = 2014,
        month = nov,
       volume = {445},
       number = {1},
        pages = {479-499},
          doi = {10.1093/mnras/stu1715},
archivePrefix = {arXiv},
       eprint = {1408.6993},
 primaryClass = {astro-ph.EP},
       adsurl = {https://ui.adsabs.harvard.edu/abs/2014MNRAS.445..479C}
}

@ARTICLE{Ida+2020,
       author = {{Ida}, Shigeru and {Muto}, Takayuki and {Matsumura}, Soko and {Brasser}, Ramon},
        title = "{A new and simple prescription for planet orbital migration and eccentricity damping by planet-disc interactions based on dynamical friction}",
      journal = {\mnras},
         year = 2020,
        month = jun,
       volume = {494},
       number = {4},
        pages = {5666-5674},
          doi = {10.1093/mnras/staa1073},
archivePrefix = {arXiv},
       eprint = {2004.07481},
 primaryClass = {astro-ph.EP},
       adsurl = {https://ui.adsabs.harvard.edu/abs/2020MNRAS.494.5666I}
}

@INPROCEEDINGS{Offner+2023,
       author = {{Offner}, S.~S.~R. and {Moe}, M. and {Kratter}, K.~M. and {Sadavoy}, S.~I. and {Jensen}, E.~L.~N. and {Tobin}, J.~J.},
        title = "{The Origin and Evolution of Multiple Star Systems}",
    booktitle = {Protostars and Planets VII},
         year = 2023,
       editor = {{Inutsuka}, S. and {Aikawa}, Y. and {Muto}, T. and {Tomida}, K. and {Tamura}, M.},
       series = {Astronomical Society of the Pacific Conference Series},
       volume = {534},
        month = jul,
        pages = {275},
          doi = {10.48550/arXiv.2203.10066},
archivePrefix = {arXiv},
       eprint = {2203.10066},
 primaryClass = {astro-ph.SR},
       adsurl = {https://ui.adsabs.harvard.edu/abs/2023ASPC..534..275O}
}

@ARTICLE{Pierens_Nelson_2007,
       author = {{Pierens}, A. and {Nelson}, R.~P.},
        title = "{On the migration of protoplanets embedded in circumbinary disks}",
      journal = {\aap},
         year = 2007,
        month = sep,
       volume = {472},
       number = {3},
        pages = {993-1001},
          doi = {10.1051/0004-6361:20077659},
archivePrefix = {arXiv},
       eprint = {0707.2677},
 primaryClass = {astro-ph},
       adsurl = {https://ui.adsabs.harvard.edu/abs/2007A&A...472..993P}
}

@ARTICLE{Sutherland_Kratter_2019,
       author = {{Sutherland}, Adam P. and {Kratter}, Kaitlin M.},
        title = "{Instabilities in multiplanet circumbinary systems}",
      journal = {\mnras},
         year = 2019,
        month = aug,
       volume = {487},
       number = {3},
        pages = {3288-3304},
          doi = {10.1093/mnras/stz1503},
archivePrefix = {arXiv},
       eprint = {1905.12638},
 primaryClass = {astro-ph.EP},
       adsurl = {https://ui.adsabs.harvard.edu/abs/2019MNRAS.487.3288S}
}

@INPROCEEDINGS{Zoppetti_2019,
       author = {{Zoppetti}, F.~A. and {Beaug{\'e}}, C. and {Leiva}, A.~M.},
        title = "{The role of the mean and forced eccentricities in the secular dynamic of circumbinary planets}",
    booktitle = {Journal of Physics Conference Series},
         year = 2019,
       series = {Journal of Physics Conference Series},
       volume = {1365},
        month = oct,
    publisher = {IOP},
          eid = {012029},
        pages = {012029},
          doi = {10.1088/1742-6596/1365/1/012029},
       adsurl = {https://ui.adsabs.harvard.edu/abs/2019JPhCS1365a2029Z}
}

@ARTICLE{Gallardo_2021,
       author = {{Gallardo}, Tabar{\'e} and {Beaug{\'e}}, Cristi{\'a}n and {Giuppone}, Cristian A.},
        title = "{Semianalytical model for planetary resonances. Application to planets around single and binary stars}",
      journal = {\aap},
         year = 2021,
        month = feb,
       volume = {646},
          eid = {A148},
        pages = {A148},
          doi = {10.1051/0004-6361/202039764},
archivePrefix = {arXiv},
       eprint = {2012.12296},
 primaryClass = {astro-ph.EP},
       adsurl = {https://ui.adsabs.harvard.edu/abs/2021A&A...646A.148G}
}

@ARTICLE{Liu_2022,
       author = {{Liu}, Beibei and {Raymond}, Sean N. and {Jacobson}, Seth A.},
        title = "{Early Solar System instability triggered by dispersal of the gaseous disk}",
      journal = {\nat},
         year = 2022,
        month = apr,
       volume = {604},
       number = {7907},
        pages = {643-646},
          doi = {10.1038/s41586-022-04535-1},
archivePrefix = {arXiv},
       eprint = {2205.02026},
 primaryClass = {astro-ph.EP},
       adsurl = {https://ui.adsabs.harvard.edu/abs/2022Natur.604..643L}
}

@ARTICLE{Orosz_2012,
       author = {{Orosz}, Jerome A. and {Welsh}, William F. and {Carter}, Joshua A. and {Fabrycky}, Daniel C. and {Cochran}, William D. and {Endl}, Michael and {Ford}, Eric B. and {Haghighipour}, Nader and {MacQueen}, Phillip J. and {Mazeh}, Tsevi and {Sanchis-Ojeda}, Roberto and {Short}, Donald R. and {Torres}, Guillermo and {Agol}, Eric and {Buchhave}, Lars A. and {Doyle}, Laurance R. and {Isaacson}, Howard and {Lissauer}, Jack J. and {Marcy}, Geoffrey W. and {Shporer}, Avi and {Windmiller}, Gur and {Barclay}, Thomas and {Boss}, Alan P. and {Clarke}, Bruce D. and {Fortney}, Jonathan and {Geary}, John C. and {Holman}, Matthew J. and {Huber}, Daniel and {Jenkins}, Jon M. and {Kinemuchi}, Karen and {Kruse}, Ethan and {Ragozzine}, Darin and {Sasselov}, Dimitar and {Still}, Martin and {Tenenbaum}, Peter and {Uddin}, Kamal and {Winn}, Joshua N. and {Koch}, David G. and {Borucki}, William J.},
        title = "{Kepler-47: A Transiting Circumbinary Multiplanet System}",
      journal = {Science},
         year = 2012,
        month = sep,
       volume = {337},
       number = {6101},
        pages = {1511},
          doi = {10.1126/science.1228380},
archivePrefix = {arXiv},
       eprint = {1208.5489},
 primaryClass = {astro-ph.SR},
       adsurl = {https://ui.adsabs.harvard.edu/abs/2012Sci...337.1511O}
}

@ARTICLE{Kostov_2020,
       author = {{Kostov}, Veselin B. and {Orosz}, Jerome A. and {Feinstein}, Adina D. and {Welsh}, William F. and {Cukier}, Wolf and {Haghighipour}, Nader and {Quarles}, Billy and {Martin}, David V. and {Montet}, Benjamin T. and {Torres}, Guillermo and {Triaud}, Amaury H.~M.~J. and {Barclay}, Thomas and {Boyd}, Patricia and {Briceno}, Cesar and {Cameron}, Andrew Collier and {Correia}, Alexandre C.~M. and {Gilbert}, Emily A. and {Gill}, Samuel and {Gillon}, Micha{\"e}l and {Haqq-Misra}, Jacob and {Hellier}, Coel and {Dressing}, Courtney and {Fabrycky}, Daniel C. and {Furesz}, Gabor and {Jenkins}, Jon M. and {Kane}, Stephen R. and {Kopparapu}, Ravi and {Hod{\v{z}}i{\'c}}, Vedad Kunovac and {Latham}, David W. and {Law}, Nicholas and {Levine}, Alan M. and {Li}, Gongjie and {Lintott}, Chris and {Lissauer}, Jack J. and {Mann}, Andrew W. and {Mazeh}, Tsevi and {Mardling}, Rosemary and {Maxted}, Pierre F.~L. and {Eisner}, Nora and {Pepe}, Francesco and {Pepper}, Joshua and {Pollacco}, Don and {Quinn}, Samuel N. and {Quintana}, Elisa V. and {Rowe}, Jason F. and {Ricker}, George and {Rose}, Mark E. and {Seager}, S. and {Santerne}, Alexandre and {S{\'e}gransan}, Damien and {Short}, Donald R. and {Smith}, Jeffrey C. and {Standing}, Matthew R. and {Tokovinin}, Andrei and {Trifonov}, Trifon and {Turner}, Oliver and {Twicken}, Joseph D. and {Udry}, St{\'e}phane and {Vanderspek}, Roland and {Winn}, Joshua N. and {Wolf}, Eric T. and {Ziegler}, Carl and {Ansorge}, Peter and {Barnet}, Frank and {Bergeron}, Joel and {Huten}, Marc and {Pappa}, Giuseppe and {van der Straeten}, Timo},
        title = "{TOI-1338: TESS' First Transiting Circumbinary Planet}",
      journal = {\aj},
         year = 2020,
        month = jun,
       volume = {159},
       number = {6},
          eid = {253},
        pages = {253},
          doi = {10.3847/1538-3881/ab8a48},
archivePrefix = {arXiv},
       eprint = {2004.07783},
 primaryClass = {astro-ph.EP},
       adsurl = {https://ui.adsabs.harvard.edu/abs/2020AJ....159..253K}
}

@ARTICLE{Zoppetti_2020,
       author = {{Zoppetti}, F.~A. and {Leiva}, A.~M. and {Beaug{\'e}}, C.},
        title = "{Tidal evolution of circumbinary systems with arbitrary eccentricities: applications for Kepler systems}",
      journal = {\aap},
         year = 2020,
        month = feb,
       volume = {634},
          eid = {A12},
        pages = {A12},
          doi = {10.1051/0004-6361/201937248},
archivePrefix = {arXiv},
       eprint = {1912.08728},
 primaryClass = {astro-ph.EP},
       adsurl = {https://ui.adsabs.harvard.edu/abs/2020A&A...634A..12Z}
}

@ARTICLE{Marzari_Thebault_2019,
       author = {{Marzari}, Francesco and {Thebault}, Philippe},
        title = "{Planets in Binaries: Formation and Dynamical Evolution}",
      journal = {Galaxies},
         year = 2019,
        month = oct,
       volume = {7},
       number = {4},
          eid = {84},
        pages = {84},
          doi = {10.3390/galaxies7040084},
archivePrefix = {arXiv},
       eprint = {2002.12006},
 primaryClass = {astro-ph.EP},
       adsurl = {https://ui.adsabs.harvard.edu/abs/2019Galax...7...84M}
}

@BOOK{Armitage_2010,
       author = {{Armitage}, Philip J.},
        title = "{Astrophysics of Planet Formation}",
         year = 2010,
       adsurl = {https://ui.adsabs.harvard.edu/abs/2010apf..book.....A}
}

@ARTICLE{Armstrong_2024,
       author = {{Armstrong}, D.~J. and {Osborn}, H.~P. and {Brown}, D.~J.~A. and {Faedi}, F. and {G{\'o}mez Maqueo Chew}, Y. and {Martin}, D.~V. and {Pollacco}, D. and {Udry}, S.},
        title = "{On the abundance of circumbinary planets}",
      journal = {\mnras},
         year = 2014,
        month = oct,
       volume = {444},
       number = {2},
        pages = {1873-1883},
          doi = {10.1093/mnras/stu1570},
archivePrefix = {arXiv},
       eprint = {1404.5617},
 primaryClass = {astro-ph.EP},
       adsurl = {https://ui.adsabs.harvard.edu/abs/2014MNRAS.444.1873A}
}

@ARTICLE{Foucart_Lai_2013,
       author = {{Foucart}, Francois and {Lai}, Dong},
        title = "{Assembly of Protoplanetary Disks and Inclinations of Circumbinary Planets}",
      journal = {\apj},
         year = 2013,
        month = feb,
       volume = {764},
       number = {1},
          eid = {106},
        pages = {106},
          doi = {10.1088/0004-637X/764/1/106},
archivePrefix = {arXiv},
       eprint = {1211.3721},
 primaryClass = {astro-ph.EP},
       adsurl = {https://ui.adsabs.harvard.edu/abs/2013ApJ...764..106F}
}

@ARTICLE{Kostov_2016,
       author = {{Kostov}, Veselin B. and {Orosz}, Jerome A. and {Welsh}, William F. and {Doyle}, Laurance R. and {Fabrycky}, Daniel C. and {Haghighipour}, Nader and {Quarles}, Billy and {Short}, Donald R. and {Cochran}, William D. and {Endl}, Michael and {Ford}, Eric B. and {Gregorio}, Joao and {Hinse}, Tobias C. and {Isaacson}, Howard and {Jenkins}, Jon M. and {Jensen}, Eric L.~N. and {Kane}, Stephen and {Kull}, Ilya and {Latham}, David W. and {Lissauer}, Jack J. and {Marcy}, Geoffrey W. and {Mazeh}, Tsevi and {M{\"u}ller}, Tobias W.~A. and {Pepper}, Joshua and {Quinn}, Samuel N. and {Ragozzine}, Darin and {Shporer}, Avi and {Steffen}, Jason H. and {Torres}, Guillermo and {Windmiller}, Gur and {Borucki}, William J.},
        title = "{Kepler-1647b: The Largest and Longest-period Kepler Transiting Circumbinary Planet}",
      journal = {\apj},
         year = 2016,
        month = aug,
       volume = {827},
       number = {1},
          eid = {86},
        pages = {86},
          doi = {10.3847/0004-637X/827/1/86},
archivePrefix = {arXiv},
       eprint = {1512.00189},
 primaryClass = {astro-ph.EP},
       adsurl = {https://ui.adsabs.harvard.edu/abs/2016ApJ...827...86K}
}

@ARTICLE{Moriwaki_Nakagawa_2004,
       author = {{Moriwaki}, Kazumasa and {Nakagawa}, Yoshitsugu},
        title = "{A Planetesimal Accretion Zone in a Circumbinary Disk}",
      journal = {\apj},
         year = 2004,
        month = jul,
       volume = {609},
       number = {2},
        pages = {1065-1070},
          doi = {10.1086/421342},
       adsurl = {https://ui.adsabs.harvard.edu/abs/2004ApJ...609.1065M}
}

@ARTICLE{Tanaka_2002,
       author = {{Tanaka}, Hidekazu and {Takeuchi}, Taku and {Ward}, William R.},
        title = "{Three-Dimensional Interaction between a Planet and an Isothermal Gaseous Disk. I. Corotation and Lindblad Torques and Planet Migration}",
      journal = {\apj},
         year = 2002,
        month = feb,
       volume = {565},
       number = {2},
        pages = {1257-1274},
          doi = {10.1086/324713},
       adsurl = {https://ui.adsabs.harvard.edu/abs/2002ApJ...565.1257T}
}

@INCOLLECTION{Martin_2018,
       author = {{Martin}, David V.},
        title = "{Populations of Planets in Multiple Star Systems}",
    booktitle = {Handbook of Exoplanets},
         year = 2018,
       editor = {{Deeg}, Hans J. and {Belmonte}, Juan Antonio},
          eid = {156},
        pages = {156},
          doi = {10.1007/978-3-319-55333-7_156},
       adsurl = {https://ui.adsabs.harvard.edu/abs/2018haex.bookE.156M}
}

@ARTICLE{Peale_1976,
       author = {{Peale}, S.~J.},
        title = "{Orbital resonance in the solar system.}",
      journal = {\araa},
         year = 1976,
        month = jan,
       volume = {14},
        pages = {215-246},
          doi = {10.1146/annurev.aa.14.090176.001243},
       adsurl = {https://ui.adsabs.harvard.edu/abs/1976ARA&A..14..215P}
}

@ARTICLE{Goldreich_Schlichting_2014,
       author = {{Goldreich}, Peter and {Schlichting}, Hilke E.},
        title = "{Overstable Librations can Account for the Paucity of Mean Motion Resonances among Exoplanet Pairs}",
      journal = {\aj},
         year = 2014,
        month = feb,
       volume = {147},
       number = {2},
          eid = {32},
        pages = {32},
          doi = {10.1088/0004-6256/147/2/32},
archivePrefix = {arXiv},
       eprint = {1308.4688},
 primaryClass = {astro-ph.EP},
       adsurl = {https://ui.adsabs.harvard.edu/abs/2014AJ....147...32G}
}

@ARTICLE{Miranda_Lai_2015,
       author = {{Miranda}, Ryan and {Lai}, Dong},
        title = "{Tidal truncation of inclined circumstellar and circumbinary discs in young stellar binaries}",
      journal = {\mnras},
         year = 2015,
        month = sep,
       volume = {452},
       number = {3},
        pages = {2396-2409},
          doi = {10.1093/mnras/stv1450},
archivePrefix = {arXiv},
       eprint = {1504.02917},
 primaryClass = {astro-ph.EP},
       adsurl = {https://ui.adsabs.harvard.edu/abs/2015MNRAS.452.2396M}
}

@ARTICLE{2022MNRAS.514.3844C,
       author = {{Charalambous}, C. and {Teyssandier}, J. and {Libert}, A.-S.},
        title = "{Proximity of exoplanets to first-order mean-motion resonances}",
      journal = {\mnras},
         year = 2022,
        month = aug,
       volume = {514},
       number = {3},
        pages = {3844-3856},
          doi = {10.1093/mnras/stac1554},
archivePrefix = {arXiv},
       eprint = {2206.00943},
 primaryClass = {astro-ph.EP},
       adsurl = {https://ui.adsabs.harvard.edu/abs/2022MNRAS.514.3844C}
}

@ARTICLE{Lee_Peale_2002,
       author = {{Lee}, Man Hoi and {Peale}, S.~J.},
        title = "{Dynamics and Origin of the 2:1 Orbital Resonances of the GJ 876 Planets}",
      journal = {\apj},
         year = 2002,
        month = mar,
       volume = {567},
       number = {1},
        pages = {596-609},
          doi = {10.1086/338504},
       adsurl = {https://ui.adsabs.harvard.edu/abs/2002ApJ...567..596L}
}

@ARTICLE{Kley_Peitz_Bryden_2004,
       author = {{Kley}, W. and {Peitz}, J. and {Bryden}, G.},
        title = "{Evolution of planetary systems in resonance}",
      journal = {\aap},
         year = 2004,
        month = feb,
       volume = {414},
        pages = {735-747},
          doi = {10.1051/0004-6361:20031589},
archivePrefix = {arXiv},
       eprint = {astro-ph/0310321},
 primaryClass = {astro-ph},
       adsurl = {https://ui.adsabs.harvard.edu/abs/2004A&A...414..735K}
}

@ARTICLE{1993Icar..103..301B,
       author = {{Beauge}, C. and {Ferraz-Mello}, S.},
        title = "{Resonance Trapping in the Primordial Solar Nebula: The Case of a Stokes Drag Dissipation}",
      journal = {\icarus},
         year = 1993,
        month = jun,
       volume = {103},
       number = {2},
        pages = {301-318},
          doi = {10.1006/icar.1993.1072},
       adsurl = {https://ui.adsabs.harvard.edu/abs/1993Icar..103..301B}
}

@ARTICLE{Orosz_2019,
       author = {{Orosz}, Jerome A. and {Welsh}, William F. and {Haghighipour}, Nader and {Quarles}, Billy and {Short}, Donald R. and {Mills}, Sean M. and {Satyal}, Suman and {Torres}, Guillermo and {Agol}, Eric and {Fabrycky}, Daniel C. and {Jontof-Hutter}, Daniel and {Windmiller}, Gur and {M{\"u}ller}, Tobias W.~A. and {Hinse}, Tobias C. and {Cochran}, William D. and {Endl}, Michael and {Ford}, Eric B. and {Mazeh}, Tsevi and {Lissauer}, Jack J.},
        title = "{Discovery of a Third Transiting Planet in the Kepler-47 Circumbinary System}",
      journal = {\aj},
         year = 2019,
        month = may,
       volume = {157},
       number = {5},
          eid = {174},
        pages = {174},
          doi = {10.3847/1538-3881/ab0ca0},
archivePrefix = {arXiv},
       eprint = {1904.07255},
 primaryClass = {astro-ph.EP},
       adsurl = {https://ui.adsabs.harvard.edu/abs/2019AJ....157..174O}
}

@ARTICLE{Nelson_2003,
       author = {{Nelson}, Richard P.},
        title = "{On the evolution of giant protoplanets forming in circumbinary discs}",
      journal = {\mnras},
         year = 2003,
        month = oct,
       volume = {345},
       number = {1},
        pages = {233-242},
          doi = {10.1046/j.1365-8711.2003.06929.x},
       adsurl = {https://ui.adsabs.harvard.edu/abs/2003MNRAS.345..233N}
}

@ARTICLE{Tanaka_Ward_2004,
       author = {{Tanaka}, Hidekazu and {Ward}, William R.},
        title = "{Three-dimensional Interaction between a Planet and an Isothermal Gaseous Disk. II. Eccentricity Waves and Bending Waves}",
      journal = {\apj},
         year = 2004,
        month = feb,
       volume = {602},
       number = {1},
        pages = {388-395},
          doi = {10.1086/380992},
       adsurl = {https://ui.adsabs.harvard.edu/abs/2004ApJ...602..388T}
}

@ARTICLE{Beauge_2003,
       author = {{Beaug{\'e}}, C. and {Ferraz-Mello}, S. and {Michtchenko}, T.~A.},
        title = "{Extrasolar Planets in Mean-Motion Resonance: Apses Alignment and Asymmetric Stationary Solutions}",
      journal = {\apj},
         year = 2003,
        month = aug,
       volume = {593},
       number = {2},
        pages = {1124-1133},
          doi = {10.1086/376568},
archivePrefix = {arXiv},
       eprint = {astro-ph/0210577},
 primaryClass = {astro-ph},
       adsurl = {https://ui.adsabs.harvard.edu/abs/2003ApJ...593.1124B}
}

@ARTICLE{Chiang_Murray-Clay_2004,
       author = {{Chiang}, Eugene I. and {Murray-Clay}, Ruth A.},
        title = "{The Circumbinary Ring of KH 15D}",
      journal = {\apj},
         year = 2004,
        month = jun,
       volume = {607},
       number = {2},
        pages = {913-920},
          doi = {10.1086/383522},
archivePrefix = {arXiv},
       eprint = {astro-ph/0312515},
 primaryClass = {astro-ph},
       adsurl = {https://ui.adsabs.harvard.edu/abs/2004ApJ...607..913C}
}

@ARTICLE{Kennedy_2019,
       author = {{Kennedy}, Grant M. and {Matr{\`a}}, Luca and {Facchini}, Stefano and {Milli}, Julien and {Pani{\'c}}, Olja and {Price}, Daniel and {Wilner}, David J. and {Wyatt}, Mark C. and {Yelverton}, Ben M.},
        title = "{A circumbinary protoplanetary disk in a polar configuration}",
      journal = {Nature Astronomy},
         year = 2019,
        month = jan,
       volume = {3},
        pages = {230-235},
          doi = {10.1038/s41550-018-0667-x},
       adsurl = {https://ui.adsabs.harvard.edu/abs/2019NatAs...3..230K}
}

@ARTICLE{Hirsh_2020,
       author = {{Hirsh}, Kieran and {Price}, Daniel J. and {Gonzalez}, Jean-Fran{\c{c}}ois and {Ubeira-Gabellini}, M. Giulia and {Ragusa}, Enrico},
        title = "{On the cavity size in circumbinary discs}",
      journal = {\mnras},
         year = 2020,
        month = oct,
       volume = {498},
       number = {2},
        pages = {2936-2947},
          doi = {10.1093/mnras/staa2536},
archivePrefix = {arXiv},
       eprint = {2008.08008},
 primaryClass = {astro-ph.EP},
       adsurl = {https://ui.adsabs.harvard.edu/abs/2020MNRAS.498.2936H}
}

@ARTICLE{Ragusa_2020,
       author = {{Ragusa}, Enrico and {Alexander}, Richard and {Calcino}, Josh and {Hirsh}, Kieran and {Price}, Daniel J.},
        title = "{The evolution of large cavities and disc eccentricity in circumbinary discs}",
      journal = {\mnras},
         year = 2020,
        month = dec,
       volume = {499},
       number = {3},
        pages = {3362-3380},
          doi = {10.1093/mnras/staa2954},
archivePrefix = {arXiv},
       eprint = {2009.10738},
 primaryClass = {astro-ph.EP},
       adsurl = {https://ui.adsabs.harvard.edu/abs/2020MNRAS.499.3362R}
}

\begin{appendix}

\section{Resonant and secular angles in the simulation described in Figure \ref{fig:Sim_qB0.6_eB0.5_simult}}
\label{app:resonant_secular_angles}

\begin{figure}[ht!]
  \centering
  \begin{overpic}[width=0.45\textwidth]{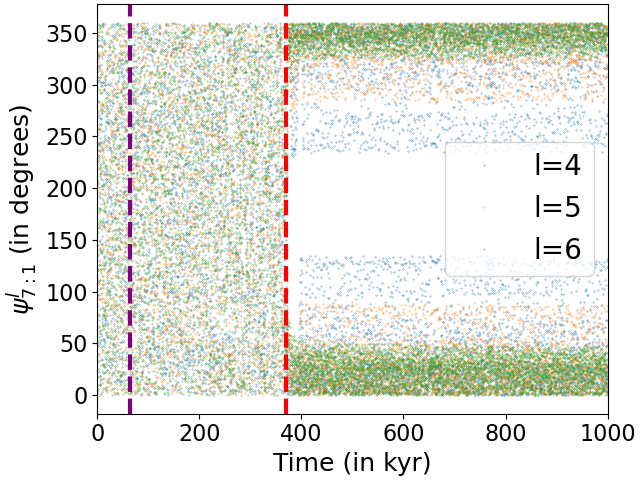}
  \put(80,57){\bfseries (a)}
  \end{overpic}
  \hfill
  \begin{overpic}[width=0.45\textwidth]{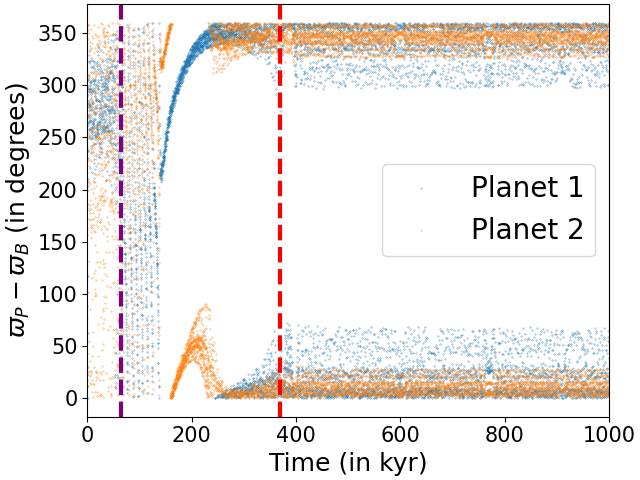}
  \put(80,57){\bfseries (b)}
  \end{overpic}
  \caption{
    Panel~(a): several resonant angles for the 7:1 MMR between Planet~1 and the binary in the simulation shown in Figure~\ref{fig:Sim_qB0.6_eB0.5_simult}. Only angles exhibiting libration are displayed: $\psi_{7:1}^{4}$, $\psi_{7:1}^{5}$, and $\psi_{7:1}^{6}$. 
    Panel~(b): temporal evolution of $\varpi_P - \varpi_B$ for both planets in the same simulation.
    In both panels, the purple vertical dashed line represents the moment when both planets enter MMR, and the red vertical dashed line represents the time when Planet~1 enters a 7:1 resonance with the binary.
  }

  \label{app:angles_combined}
\end{figure}

Planet~1 in the simulation shown in Fig.~\ref{fig:Sim_qB0.6_eB0.5_simult} (with $q_B = 0.6$ and $e_B = 0.5$ in the simultaneous scenario) enters a 7:1 MMR with the binary at $t = 370$ kyr. This is evidenced in Panel~(a) of Fig.~\ref{app:angles_combined}, where several resonant angles defined in Eq.~\ref{eq:resonant_angles} start librating around 0° after this time. In addition, both planets enter a secular resonance shortly after becoming mutually resonant at $t = 65$ kyr, as shown in Panel~(b) of Fig.~\ref{app:angles_combined}, where the angles $\varpi_{P} - \varpi_{B}$ cease to circulate around $t = 140$ kyr. Systems that are simultaneously in mean-motion and secular resonance are expected to evolve near Apsidal Corotation Resonance (ACR) solutions, corresponding to stationary configurations of the averaged equations of motion. \cite{Beauge_2003} studied such solutions for two circumstellar planets in 2:1 and 3:1 MMRs and identified several families, including aligned configurations in which both resonant and secular angles librate around 0°. By analogy, our results suggest that Planet~1 and the binary evolve close to an aligned ACR configuration in this simulation.

\FloatBarrier
\section{Residual maps of the differences between the two migration scenarios}

\begin{figure}[ht!]
\centering
    \begin{tabular}{c}
        \begin{overpic}[width=0.45\textwidth]{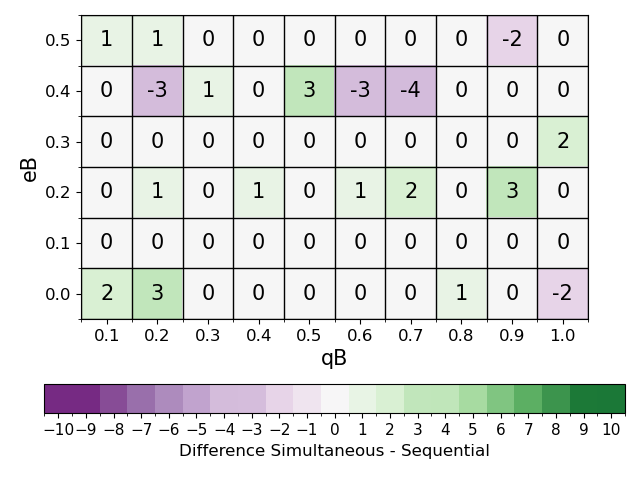}
        \put(50,-1){\bfseries (a)}
        \end{overpic} \\
        \begin{overpic}[width=0.45\textwidth]{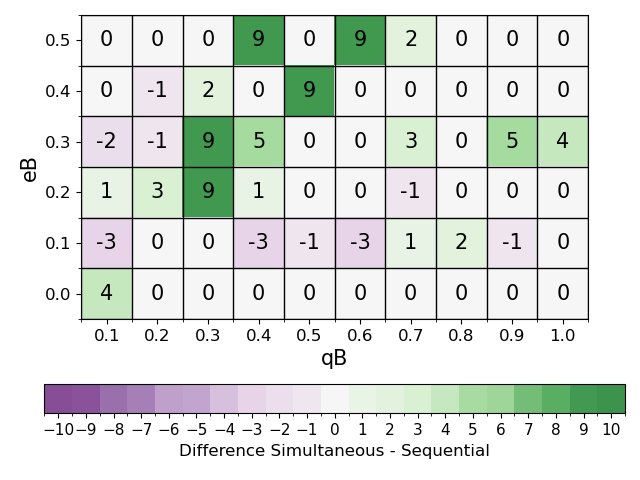}
        \put(50,-1){\bfseries (b)}
        \end{overpic}
    \end{tabular}
    
    \caption{
        Residual maps obtained by making the difference between stability maps for the simultaneous scenario and those for the sequential scenario. Panel~(a): residual map for systems with at least one planet remaining. Panel~(b): residual map for systems keeping both planets stable. The x-axis (resp. the y-axis) indicates the binary mass ratio (resp. binary eccentricity) studied.
    }
    \label{fig:stabilitymap_diff}
    
\end{figure}

\FloatBarrier
\clearpage
\section{Methodology to determine the MMR of a planet by using the complex number $z_l$}
\label{app:complex_number_zl}

For a given resonance \textit{p:q} under consideration, we define the complex number $z_l = \exp(-j.\psi_{p:q}^l)$, where $\psi_{p:q}^{l}$ denotes the $l$-th resonant angle defined in Equation~\ref{eq:resonant_angles}. When two bodies are not trapped in the considered resonance, the associated resonant angles are expected to circulate, spanning the full range from $0^\circ$ to $360^\circ$. As a consequence, the time average of $z_l$ over the duration of the simulation $|\langle z_l \rangle|$ is expected to approach zero. Conversely, when the two bodies are locked in resonance, the resonant angles librate around a constant value and spans only a restricted range around this constant value. Its time-averaged value becomes non-zero, with a magnitude that strongly depends on the libration amplitude: the smaller the libration amplitude, the closer $|\langle z_l \rangle|$ is to unity; conversely, large libration amplitudes lead to smaller values, approaching zero in the limit of circulation. We therefore adopt the magnitude of the time-averaged complex number, $|\langle z_l \rangle|$, as a quantitative criterion to determine whether two bodies are in MMR.

\begin{figure}[ht!]
  \centering
  \begin{minipage}{0.45\textwidth}
    \includegraphics[width=1\textwidth]{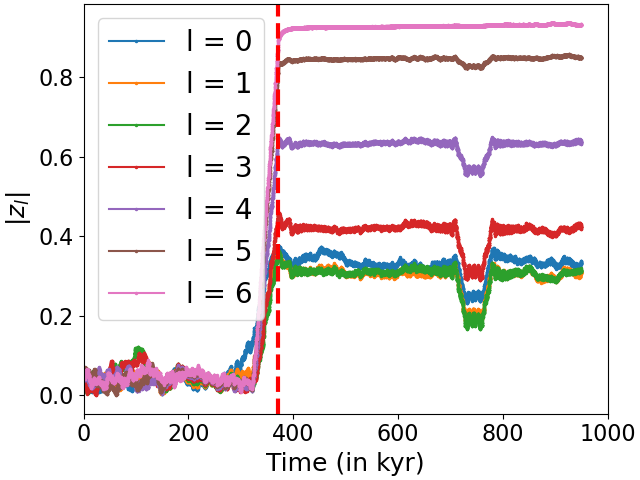}
  \end{minipage}
  \caption{\centering
    Evolution of the average value of $|\langle z_l \rangle|$ over time with a integration window of 50 kyr. The red vertical dashed line indicates the moment when Planet~1 enters a 7:1 MMR with the binary.
  }
  \label{fig:mean_z}
\end{figure}

However, this criterion is not strictly applicable in the context of circumbinary planets. In such systems, resonant angles are not expected to exhibit simple, regular (i.e., quasi-sinusoidal and strictly periodic) librations, but rather more complex and possibly modulated behavior, due to perturbations from the central binary and mutual planetary interactions. Consequently, the time-averaged quantity $|\langle z_l \rangle|$ may not approach unity even in true resonant configurations. To evaluate the robustness of this criterion, we examined the temporal evolution of $|\langle z_l \rangle|$ for several representative simulations in our dataset (Fig.~\ref{fig:mean_z} shows the evolution of $|\langle z_l \rangle|$ for the resonance illustrated in Fig.~\ref{fig:Sim_qB0.6_eB0.5_simult}). In all inspected cases, resonant angles exhibiting clear libration systematically correspond to values of $|\langle z_l \rangle| \gtrsim 0.5$. However, this threshold alone is not sufficient. We identify several configurations in which planets are trapped in a given resonance, but other candidate resonant angles (associated with different commensurabilities) do not librate but still yield $|\langle z_l \rangle|$ values slightly above 0.5. For example, in a configuration where a planet is in a 7:1 MMR with the binary, the resonant angles corresponding to a 4:1 commensurability do not librate, but produce $|\langle z_l \rangle| \sim 0.6$. In practice, however, the value associated with the true resonant configuration remains systematically higher than those obtained for non-librating angles. We therefore identify the actual resonance by selecting, among all tested commensurabilities, the one that maximizes $|\langle z_l \rangle|$.

The procedure applied to each simulation is as follows:

\begin{itemize}
  \item First, we verify whether at least one planet remains dynamically stable at the end of the simulation. Only simulations satisfying this condition are further analysed.
  \item Second, we define the set of MMRs to be tested. For resonances with the binary, Figure 3 of \cite{Martin_Fitzmaurice2022} indicates that, within our parameter space, we expect N:1 resonances with $3 \le N \le 9$. These resonances therefore constitute our test list.
  \item Third, we define the time window over which $|\langle z_l \rangle|$ is computed. We adopt a window of 50 kyr. This duration ensures a sufficiently large number of data points to obtain a statistically robust average, while remaining short compared to the typical duration of stable resonant phases, which generally last several hundred thousand years.
  \item Fourth, for each resonance in the test list, we compute all values of $|\langle z_l \rangle|$ over the time interval extending from (simulation end - 50 kyr) to the end of the simulation. If any value exceeds 0.5, we retain the maximum value obtained for that resonance in the so-called "potential MMRs list".
  \item Finally, if at least one resonance yields a retained maximum value, we select the largest one among those in the potential MMRs list and identify the corresponding resonance as the MMR established in the simulation.
\end{itemize}

Nevertheless, we note that this method for identifying resonances has several limitations. First, in a number of simulations, a planet ceases its migration without becoming trapped in resonance with either the binary or the innermost planet (i.e., without any libration of the resonant angles). However, some of these cases exhibit resonant-angle evolution profiles showing clusters of points around a nearly constant value. As a result, the $|\langle z_l \rangle|$ value can exceed 0.5, leading these simulations to be incorrectly classified as resonant. Consequently, this effect tends to overestimate the number of resonant cases in our sample. One possible way to mitigate this issue is to increase the threshold above which a simulation is considered resonant. However, doing so would introduce the opposite bias by underestimating the number of resonant cases, since genuine resonant configurations could be excluded by adopting an excessively restrictive threshold. Unfortunately, no single threshold value completely eliminates both over- and underestimation. Its selection must instead reflect a compromise between these two competing effects. In addition, the use of a temporal window of 50 kyr may mask a resonance disruption (or a resonant capture) occurring during the final 50 kyr of the simulation. Indeed, the computation of the last $|\langle z_l \rangle|$ value retains information from the preceding resonant phase, which can influence the final classification. Among the simulations that we inspected individually, we did not encounter such situations; in most cases, resonance breaking is followed almost immediately by the ejection of the planet. Yet, we cannot rule out the possibility that a small number of simulations have been incorrectly identified as resonant.

\FloatBarrier
\clearpage
\onecolumn
\section{Evolution of mean-motion ratios in simulations using a migration model with varying migration timescales}
\label{app:Mosaic_realistic-migration-tests}

\begin{figure*}[ht!]
  \centering
  \begin{minipage}{0.9\textwidth}
    \includegraphics[width=1\textwidth]{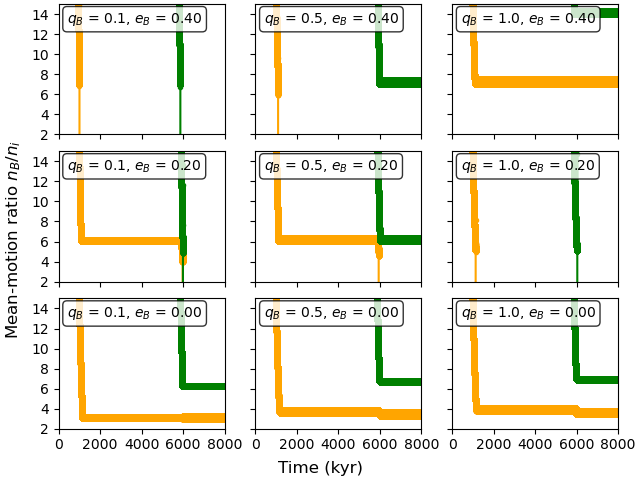}
  \end{minipage}
  \caption{\centering
    Evolution of the mean-motion ratios between the planets and the binary star in simulations adopting the migration model from \cite{Cresswell_Nelson2008}, within the framework of the sequential migration scenario. The orange curve shows the mean-motion ratio between the binary and Planet~1, while the green curve shows the mean-motion ratio between the binary and Planet~2.
  }
  \label{fig:mosaic_migration_tests}
\end{figure*}

\end{appendix}

\end{document}